\documentclass[preprint,12pt,authoryear]{elsarticle}
\usepackage{latexsym}
\usepackage{epsfig}
\usepackage{amssymb,amsmath,verbatim,mathtools,needspace,enumitem,etoolbox,graphicx,physics,microtype,ulem}
\usepackage{lineno}
\usepackage{float,xcolor}
\usepackage{verbatim}
\def\be{\begin{equation}}
\def\ee{\end{equation}}

\newcommand{\lcdm}{$\Lambda$CDM}
\newcommand{\g}{$\gamma$}
\newcommand{\Fermi}{\it{Fermi}\rm}
\newcommand{\planck}{\it{Planck}\rm}

\journal{Searching for ISW Effect from \Fermi-LAT}

\begin{document}

\begin{frontmatter}

\title{Searching for Integrated Sachs-Wolfe Effect from {\Fermi}-LAT Diffuse {\g}-ray Map}

\author[aa1,aa2,aa3,aa4,aa5]{Xiu-Hui Tan}
\ead{tanxh@ihep.ac.cn}
\author[aa5]{Ji-Ping Dai}
\ead{daijp@mail.bnu.edu.cn}

\author[aa5]{Jun-Qing Xia\corref{coau}}
\cortext[coau]{Corresponding author}
\ead{xiajq@bnu.edu.cn}
\address[aa1]{Institute of High Energy Physics, Chinese Academy of Sciences, Beijing 100049, China}
\address[aa2]{School of Physical Sciences, University of Chinese Academy of Sciences, Beijing 100049, China}
\address[aa3]{Dipartimento di Fisica, Universit\`a di Torino, Via Pietro Giuria 1, I-10125, Torino, Italy}
\address[aa4]{INFN – Istituto Nazionale di Fisica Nucleare, Sezione di Torino, via P. Giuria 1, I-10125 Torino, Italy}
\address[aa5]{Department of Astronomy, Beijing Normal University, Beijing 100875, China}

\begin{abstract}
	
In this paper, we estimate the cross-correlation power spectra between the {\planck} 2018 cosmic microwave background (CMB) temperature anisotropy map and the unresolved \g-ray background (UGRB) from the 9-years \Fermi-Large Area Telescope (LAT) data.
In this analysis, we use up to nine energy bins over a wide energy range of [0.631, 1000] GeV from the {\Fermi}-LAT UGRB data.
Firstly, we find that the {\Fermi} data with the energy ranges [1.202, 2.290] GeV and [17.38, 36.31] GeV show the positive evidence for the Integrated Sachs-Wolfe (ISW) effect at $1.8\sigma$ confidence level, and the significance would be increased to $2.7\sigma$ when using these two energy bins together.
Secondly, we apply the single power-law model to normalize the amplitude and use all the nine {\Fermi} energy bins to measure the significance of the ISW effect, we obtained ${\rm A_{amp}}=0.95 \pm 0.53$ ($68\%$ C.L.).
For the robustness test, we implement a null hypothesis by randomizing the {\Fermi} mock maps of nine energy bins and obtain the non-detection of ISW effect, which confirms that the ISW signal comes from the {\Fermi}-LAT diffuse \g-ray data and is consistent with the standard {\lcdm} model prediction essentially. We use a cross-correlation coefficient to show the relation between different energy bins.
Furthermore, we vary the cut ranges $|b|$ of galactic plane on the mask of {\Fermi} map and carefully check the consequent influence on the ISW signal detection.

\end{abstract}
\end{frontmatter}

\section{Introduction}
\label{sec:intro}

The nature of dark energy is one of the most important question in the modern cosmology.
As we know, dark energy is a predominated ingredient of the Universe, which takes 68\% of the whole constituents \citep{Aghanim:2018eyx}.
Measurements of the baryon acoustic oscillation \citep{Eisenstein:2005su}, distant Type Ia supernovae \citep{Riess:2004nr} and gravitational lensing \citep{Bacon:2000sy} give a good explanation and evidence that dark energy drives the acceleration of the Universe and the standard {\lcdm} theory can successful explain these observations very well.
Furthermore, Cosmic Microwave Background (CMB) as a landmark in Big Bang cosmology theory, is a crucial measurement for many open problems.
Besides, the CMB anisotropies which produced at very early times, have several important secondary effects, such as the CMB lensing and the Sunyaev-Zel'dovich effects, which could contribute additional anisotropies on the temperature of CMB photons.

The integrated Sachs-Wolfe effect (ISW; \citet{Sachs:1967er}) is also one of the CMB secondary anisotropies, which is caused by cumulative effect of photons travel from the changing of the gravitational potentials after the recombination epoch.
When a CMB photon falls into a gravitational potential well, it gains energy, while it loses energy when it climbs out.
These effects would be cancelled if the potential is time independent, such as the matter dominated era in which the gravitational potential stays constant.
However, when dark energy or curvature become important at later times, the potential evolves as the photon passes through it.
In this case, additional CMB anisotropies will be produced.
Therefore, observing the late-time ISW can be a powerful way to probe dark energy and its evolution.

However, the most significant ISW effect contributes to the CMB anisotropies on large scales that are strongly affected by the cosmic variance, which means it is quite challenging to directly extract the ISW information from the CMB observation.
Fortunately, this problem can be solved by cross-correlating ISW temperature fluctuation and the density of astrophysical objects like galaxies \citep{Crittenden:1995ak}.
Such cross-correlation analysis has been already implemented in the literatures to detect the ISW effect.
For the first detection, \citet{Crittenden:1995ak} and \citet{Boughn:1997vs} have measured the cross-correlation of High Energy Astronomy Observatory (HEAO) X-ray data and CMB anisotropies from Cosmic Background Explorer (COBE), they claimed it is potential that the ISW effect as an important observation to confront the structures of the Universe.
Afterwards, a similar set of analyses have been carried out which relied on CMB data from the Wilkinson Microwave Anisotropy Probe (WMAP) satellite and a variety of large scale structure (LSS) probes, such as the Two-Micron All-Sky Survey galaxy (2MASS) \citep{Afshordi:2003xu,Francis:2009pt,Dupe:2010zs}, the NRAO VLA Sky Survey (NVSS) radio galaxies \citep{Nolta:2003uy}, galaxies from Sloan Digital Sky Survey (SDSS) \citep{Scranton:2003in,Fosalba:2003ge,Granett:2008ju,Chen:2015wga}, the Wide-field Infrared Survey Explorer (WISE) survey \citep{Goto:2012yc,Ferraro:2014msa}, and the X-ray background HEAO catalogue \citep{Boughn:2002bs, Boughn:2003yz}.
Furthermore, \citet{Giannantonio:2008zi,Xia:2011ax,Khosravi:2015boa} have combined all these measurements together to detect the ISW effect with very high significance.
{Recently, based on current {\planck} data release, \cite{Ade:2013dsi} used the different methods gave the significance of detection ranges from 2 to 4$\sigma$;
the detection level achieved at 3$\sigma$ in \cite{Ade:2015dva} by combining the cross-correlation signal coming from all the galaxy catalogues;
\citet{Stolzner:2017ged} also combined various LSS tracer data sets in the radio, optical and infrared wavelength, and obtained more than $2\sigma$ detection of the ISW effect.}

In our previous works \citep{Camera:2014rja,Cuoco:2015rfa,Regis:2015zka,Xia:2015wka,Troster:2016sgf,Cuoco:2017bpv,Branchini:2016glc,Colavincenzo:2019jtj,Tan:2019gmb}, we have studied that gamma-ray maps of the Unresolved \g-ray background (UGRB) from {\Fermi}-Large Area Telescope (LAT) data and cross-correlated with different catalogs of galaxies. We observed significant correlation after properly removing various contamination, which means that the UGRB from {\Fermi} data could be a nice LSS tracer and, therefore, has the potential to search for the ISW effect.
Therefore, here we use the UGRB from \Fermi-LAT 9-year Pass 8 data release and the {\planck} 2018 temperature anisotropies to detect the ISW effect.
This paper is structured as follows: in section \ref{sec:data}, we introduce the data set we use; theoretical formulae are present in section \ref{sec:tform}; sections \ref{sec:statistic} and \ref{sec:null} give numerical results and some systematic checks; final summary is listed in section \ref{sec:concl}.

\section{Data Sets}
\label{sec:data}

In this section, we briefly describe the CMB map from {\planck} satellite and \g-ray maps of the UGRB from the \Fermi-LAT mission used in the analysis.

\subsection{CMB Information}

The CMB data we used comes from the \planck 2018 full-mission data release (``PR3") that are accessible on the Planck Legacy Archive (PLA\footnote{http://pla.esac.esa.int/pla}).
There are four methodologies (\texttt{COMMANDER}, \texttt{SEVEM}, \texttt{SMICA}, \texttt{NILC}) adopted to do the component-separation algorithms for maps and we choose the \texttt{SMICA} synthesized CMB map, which uses spectral matching technology \citep{Akrami:2018mcd}.
Since the contribution of the ISW signal is mainly from large scales, usually we only need to use the maps with relative low resolution format, like $N_{\rm side}=64$ or 128.

\begin{figure}
    \centering
    \includegraphics[width=8cm]{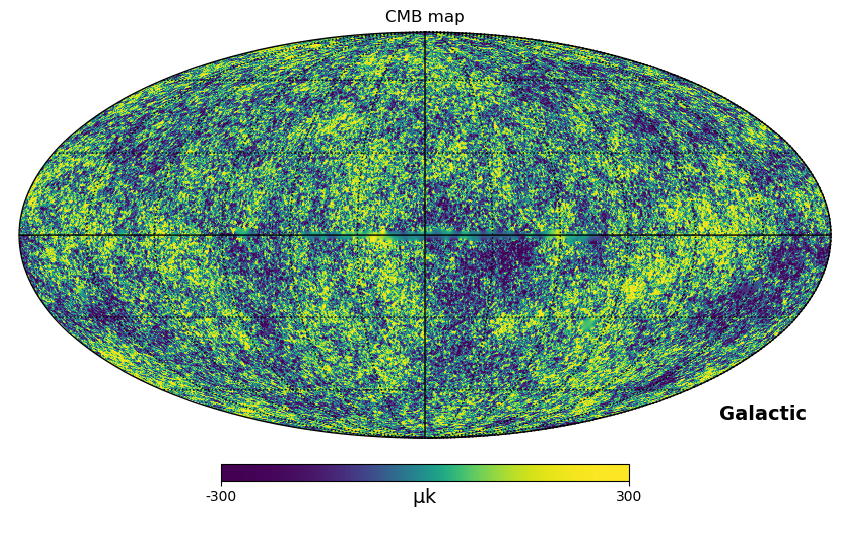}
    \includegraphics[width=8cm]{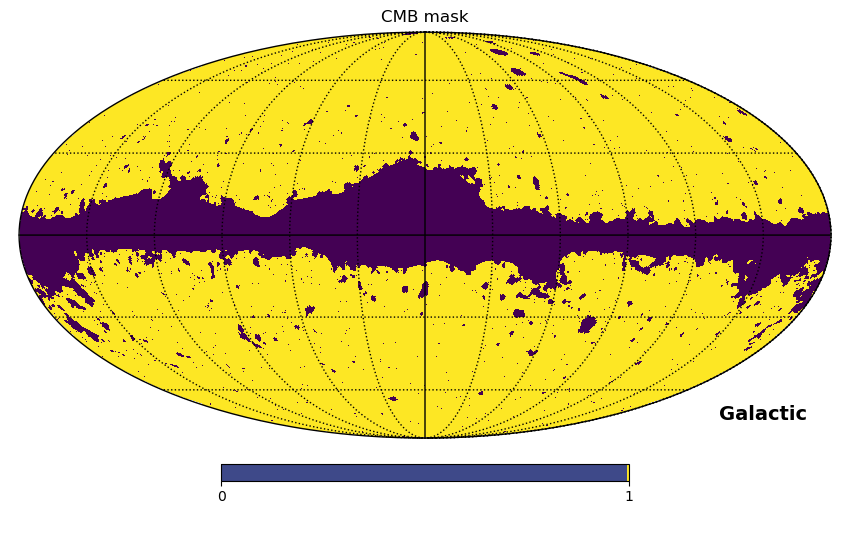}
    \caption{The CMB and mask maps used in the analysis with $N_{\rm side}=1024$.}
    \label{fig:cmbmap}
\end{figure}

In this work, in order to avoid some extra uncertainties during degradation of {\Fermi} maps, here we use the {\planck} map with $N_{\rm side}=1024$ to match the default format of {\Fermi} maps we have generated, which shown in the first panel of Fig. \ref{fig:cmbmap}.
Concerning the strong contamination from the galactic foreground and confident point-sources, in the second panel of Fig. \ref{fig:cmbmap} we also show the temperature analysis mask provided by the {\planck} team which leaves about $f_{sky}=83.8\%$ of the sky.

\subsection{\Fermi-LAT UGRB Maps}

The \Fermi-LAT is the pair-conversion \g-ray telescope with the high sensitivity to \g-rays in the high energy range from 20 MeV to 1 TeV for all-sky area.
During 10 years of operation, it has been provided massive data for researches of the astrophysical sources, cosmic rays and UGRB, and so on.
The \Fermi-LAT collaboration has also produced several generations of high-energy \g-ray source catalogues over many years, including active galactic nuclei, pulsars and other kinds of extra-galactic and galactic sources.

In this work, we use the photon flux intensity maps in 9 energy bins (see Table. \ref{tab:bins}) from 631 MeV to 1 TeV over 108 months.
The flux map is obtained from dividing the photon count maps by the exposure maps and the pixel area, and these corresponding exposure and count maps are directly produced by using the LAT Fermi tools \texttt{v10r0p5}\footnote{https://fermi.gsfc.nasa.gov/ssc/data/analysis/documentation/Cicerone/}.
The event class is selected from the Pass 8 ULTRACLEANVETO, which provides lower background contamination at low energies.
Since the observed data has an energy dependent angular resolution of the instrument, to balance the point-spread function (PSF) and photon statistics, PSF3 and PSF1+2+3 events are used for the first bin and the rest of eight bins, respectively.
The PSF is computed running the routine \texttt{gtpsf} from the Fermi-LAT Science Tools and legendre transform to window beam function to correct our analysis.

For the masks, we removed the contributions from the resolved sources list of 4FGL and 3FHL catalogs above 9 GeV.
The 4FGL catalog is updated from the \g-ray sky with respect to 3FGL and has exposure twice longer, and contains nearly 5065 sources above $4\sigma$ significance.
The 3FHL catalog, the third catalog of hard \Fermi-LAT sources, contains 1556 objects characterized in the energy range [10, 2000] GeV.

This catalog was constructed by the first 7 years of data, and is updated by implementing improvements provided by the Pass 8 data.
The radius is considered by the PSF for a gaussian method according to the containment angle.
To reduce the impact of the Galactic emission on our analysis and focused on the UGRB, we apply a Galactic latitude cut $|b|>30^{\circ}$, which represents the best compromise between pixel statistics and Galactic contamination, found in \citet{Xia:2011ax}, in order to mask the bright emission along the Galactic plane.

\begin{figure}
    \centering
    \includegraphics[width=12cm]{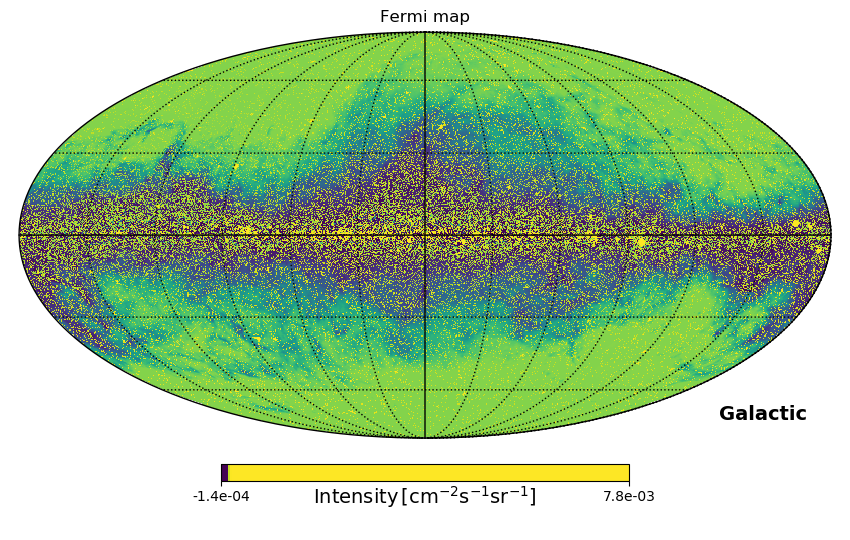}
    \caption{\Fermi-LAT flux map without the mask in the energy range [0.631, 1.202] GeV.}
    \label{fig:fermimap}
\end{figure}

All the flux maps are in the same format with the {\planck} CMB map as described above with $N_{side}=1024$, containing $N_{pix}=12,582,912$ pixels with $0.06^{\circ}$ space in each pixel.
The foreground contamination is rejected by the Galactic emission model \texttt{gll\_iem\_v06.fits}\footnote{https://fermi.gsfc.nasa.gov/ssc/data/access/lat/BackgroundModels.html}, which is provided by {\Fermi}-LAT Collaboration in same $N_{\rm side}$ number for the high-latitudes.
One example map shows in Fig. \ref{fig:fermimap},
{here we consider the \g-ray maps as continuous field on the celestial sphere, we access the density number of photons per map of one energy bin showed in table \ref{tab:bins}.}

\begin{table}
    \centering
    \caption{9 energy bins in GeV for \Fermi-LAT flux maps. $E_{\rm min}$ and $E_{\rm max}$ are the lower and upper bound of the bins, respectively. The third column is the photon number density for every energy bin maps.
    }
    \begin{tabular}{cccc}
        \hline
        \hline
        Bin & $E_{\rm min}$ [GeV] & $E_{\rm max}$  [GeV] &Photon number density \\
        &&&[ph/$\rm{deg^2}$]\\
        \hline
         1 & 0.631 & 1.202 &15.24\\
         2 & 1.202 & 2.290 &17.36\\
         3 & 2.290 & 4.786 &15.29\\
         4 & 4.786 & 9.120 &7.74\\
         5 & 9.120 & 17.38 &4.12\\
         6 & 17.38 & 36.31 &2.50\\
         7 & 36.31 & 69.18 &0.80\\
         8 & 69.18 & 131.8 &0.28\\
         9 & 131.8 & 1000 &0.037\\
         \hline
    \end{tabular}
    \label{tab:bins}
\end{table}

\section{Theoretical formalism}
\label{sec:tform}

Following \citet{Xia:2015wka}, the cross-correlation power spectrum between $\gamma$-ray maps and CMB temperature anisotropic map can be easily obtained by,
\be
C_{l}^{\gamma\rm T}=\frac{2}{\pi} \int k^{2} P(k)\left[G_{\ell}^{\gamma}(k)\right]\left[G_{\ell}^{\rm T}(k)\right] d k~,
\ee
where $k$ is the wavenumber, $P(k)$ is the matter power spectrum at present time, $G_{\ell}^{\gamma}(k)$ and $G_{\ell}^{\rm T}(k)$ are the window functions for \g-ray and CMB observations, respectively.
In the following, we derive the cross-correlation power spectrum by adopting the standard general relativity in the flat {\lcdm} framework.
While the ISW effect for non-zero curvature or modified gravity, please refer to some previous studies \citep{Kamionkowski:1996ra, Song:2006ej}.

For the temperature fluctuations from CMB maps, the ISW signal in real space can be written as
\be
\Theta(\hat{n})=-2 \int \frac{d \Phi(\hat{n} \chi, \chi)}{d \chi} d \chi~,
\ee
where $\Phi$ is the gravitational potential and $\chi$ represents the comoving distance. Here, we neglect a optical depth factor of $e^{-\tau}$, which would not affect our results because of the tiny modification when comparing with the typical accuracy achieved in the determination of the ISW effect itself.
Using the Poisson equation, we have
\be
\Phi(k, z)=-\frac{3}{2 c^{2}} \frac{\Omega_{m}}{a(z)} \frac{H_{0}^{2}}{k^{2}} \delta(k, z)~,
\ee
where $c$ is the speed of light, $a$ is the cosmological scale factor, $H_0$ is the Hubble parameter at present time, $\Omega_m$ is the fractional density of matter today, and $\delta(k,z)$ is the matter fluctuation field in Fourier space.
Finally we can obtain
\be
G_{\ell}^{\rm T}(k)=\frac{3 \Omega_{m}}{c^{2}} \frac{\mathrm{H}_{0}^{2}}{k^{2}} T_{\rm CMB}\int \frac{d}{d z}\left(\frac{D(z)}{a(z)}\right) j_{\ell}[k \chi(z)] d z~,
\ee
where $D(z)$ is the linear growth factor of density fluctuation, $j_{\ell}[k \chi(z)]$ represent spherical Bessel functions, and $T_{\rm CMB}$ is the mean temperature of the CMB photons at present.

We presume the variations that the number density of unresolved sources $n_\gamma(\hat{n} \chi, z)$ are responsible for the local fluctuation in the $\gamma$-ray luminosity density, $\rho(\hat{n} \chi, z)$.
Therefore, the two fluctuations in $n_\gamma$ and $\rho_\gamma$ are related through
\be
\frac{\rho_{\gamma}(\hat{n} \chi, z)-\rho_{\gamma}(z)}{\rho_{\gamma}(z)}=\frac{n_{\gamma}(\hat{n} \chi, z)-n_{\gamma}(z)}{n_{\gamma}(z)}.
\ee
Then, a local linear bias between the number density of objects and the underlying mass density $\rho_m$ can be regarded as,
\be
\delta_{n_{\gamma}}(\hat{n} \chi, z) \equiv b_{\gamma}(z) \delta_{m}(\hat{n} \chi, z)=b_{\gamma}(z) \frac{\rho_{m}(\hat{n} \chi, z)-\rho_{m}(z)}{\rho_{m}(z)}
\ee

where $b_\gamma(z)$ is the redshift-dependent bias parameter of the \g-ray emitters.
Finally, the contribution from the diffuse UGRB field $G_{\ell}^{\gamma}(k)$ can be expressed as,
\be
G_{\ell}^{\gamma}(k)=\int \rho_{\gamma}(z) b_{\gamma}(z) D(z) j_{\ell}[k \chi(z)] d z~,
\ee
where $\rho_{\gamma}(z)$ is the normalized luminosity density.
In our analysis, we use one astrophysical \g-ray emission source model (Star-Forming Galaxies) to interpret the UGRB information, since the significance of ISW effect is almost independent on the selected theoretical model.
We use the bias parameter and the normalized luminosity density of this type of astrophysical source provided in \citet{Xia:2015wka}.
{This bias can be computed in terms of the halo bias, depending on the \g-ray luminosity of the source and the mass of the host DM halo. Therefore it is characterised by its own bias factor of star-forming galaxy, which is the class of sources we use here, and it dose not depend on the different eneryg of {\Fermi} \g-ray maps.}

Furthermore, we also use the Limber approximation \citep{Limber:1954zz} due to the enough accuracy at scales we are interested in ($\ell>10$, see more information in section \ref{sec:statistic}), the cross correlation power spectrum can be written as
\be
\begin{aligned} C_{\ell}^{\gamma\rm T}=&\frac{3 \Omega_{m} \mathrm{H}_{0}^{2}T_{\rm CMB}}{c^{3}\left(l+\frac{1}{2}\right)^{2}} \int d z\rho_{\gamma} b_{\gamma}(z)  H(z) D(z)  \\ & \times\frac{d}{d z}\left(\frac{D(z)}{a(z)}\right) P\left(k=\frac{l+0.5}{\chi(z)}\right)B(\ell)~, \end{aligned}
\ee
here we include the correction of Gaussian beam $B(\ell)$ with the beam size $\Theta_{\rm FWHM}=5$ arcmin.
Usually the measured power spectrum will be different from the {\lcdm} prediction.
To quantify this we scale the model with a free amplitude parameter $A_{\rm amp}$: $\hat{C}_{\ell}^{\gamma\rm T} = A_{\rm amp}{C}_{\ell}^{\gamma\rm T}$, where $\hat{C}_{\ell}$ and ${C}_{\ell}$ are the measured and theoretical cross-correlation power spectra, respectively.

\begin{figure*}
	\centering
	\includegraphics[width=4.7cm]{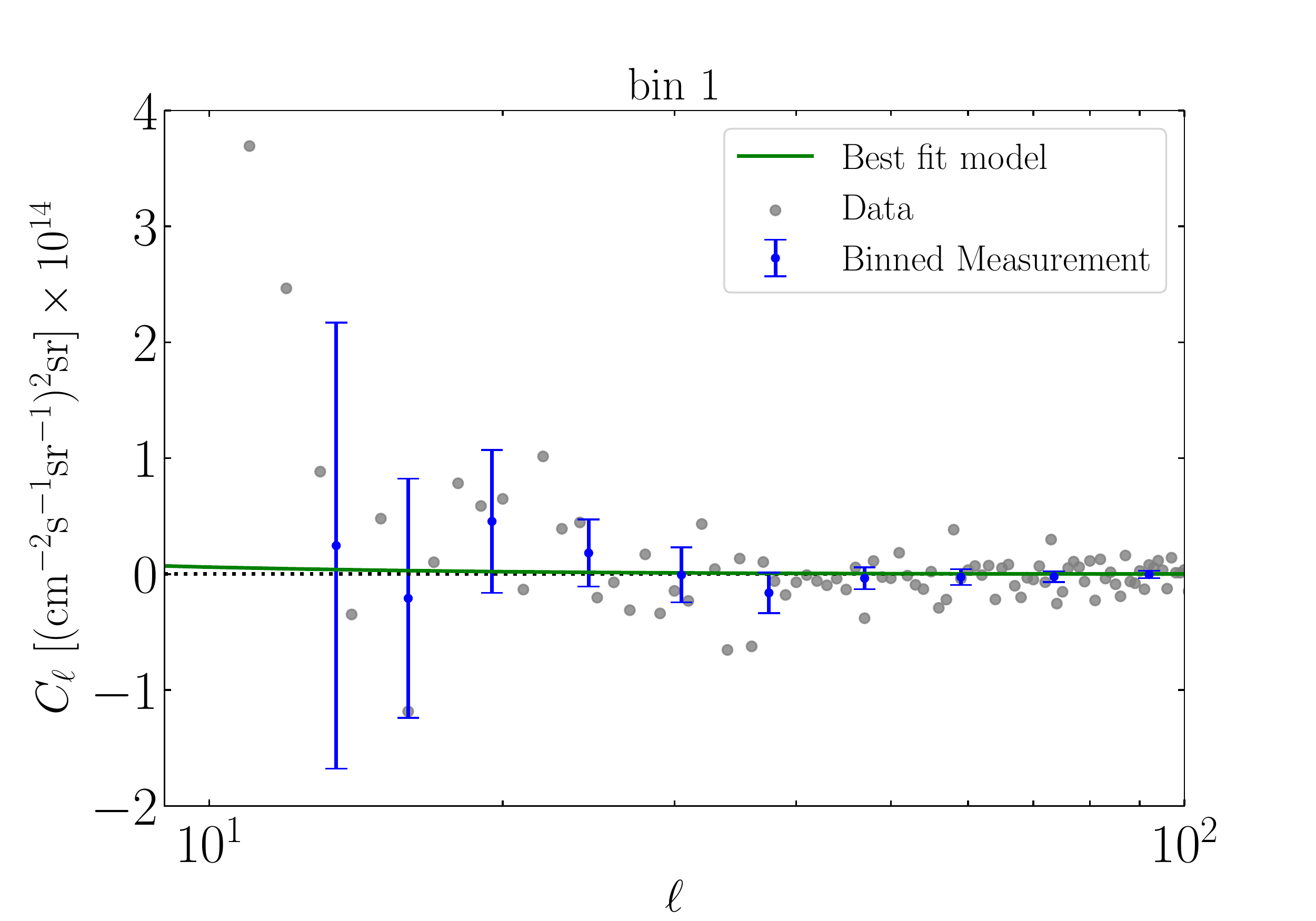}
	\includegraphics[width=4.7cm]{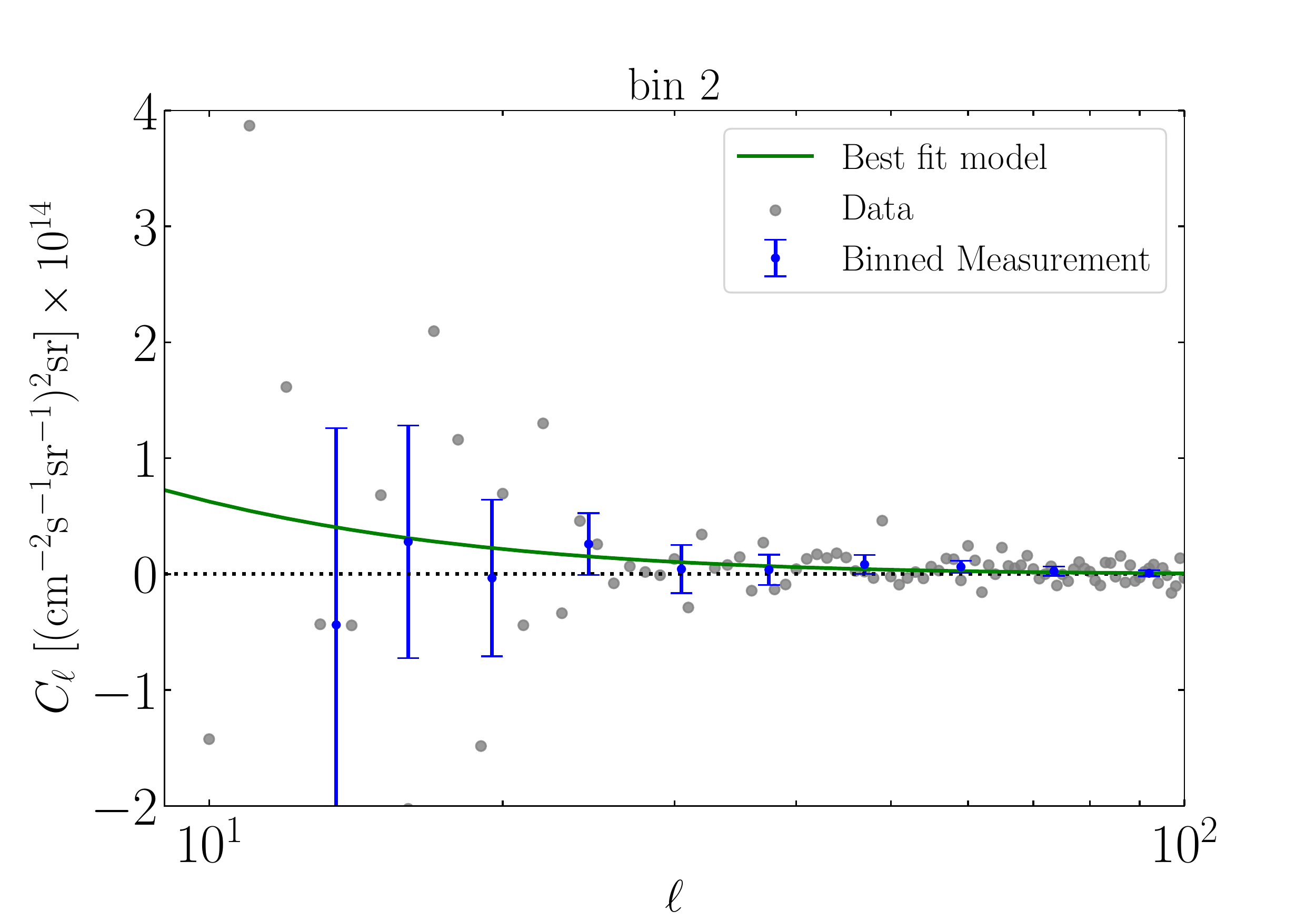}
	\includegraphics[width=4.7cm]{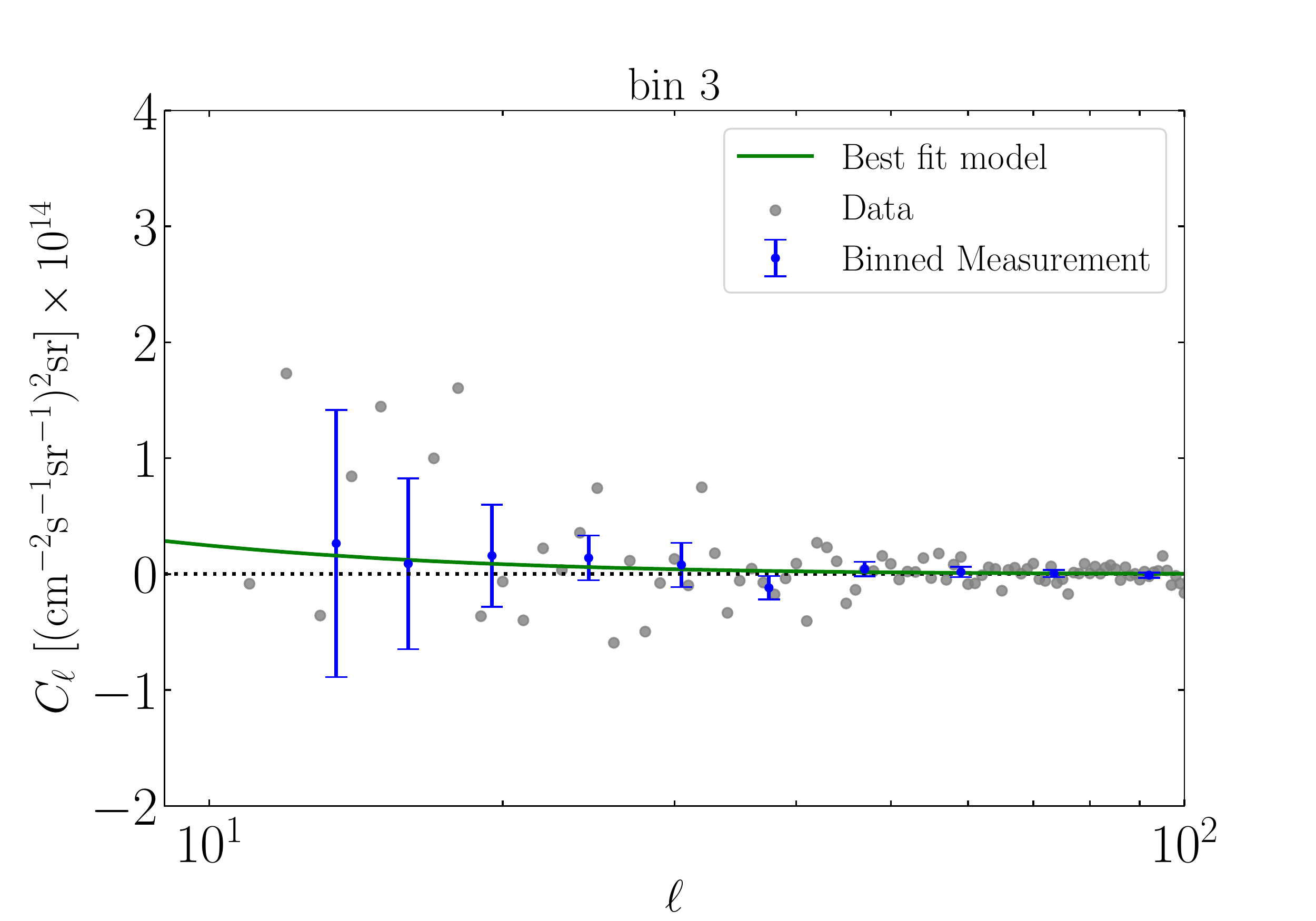}
	\includegraphics[width=4.7cm]{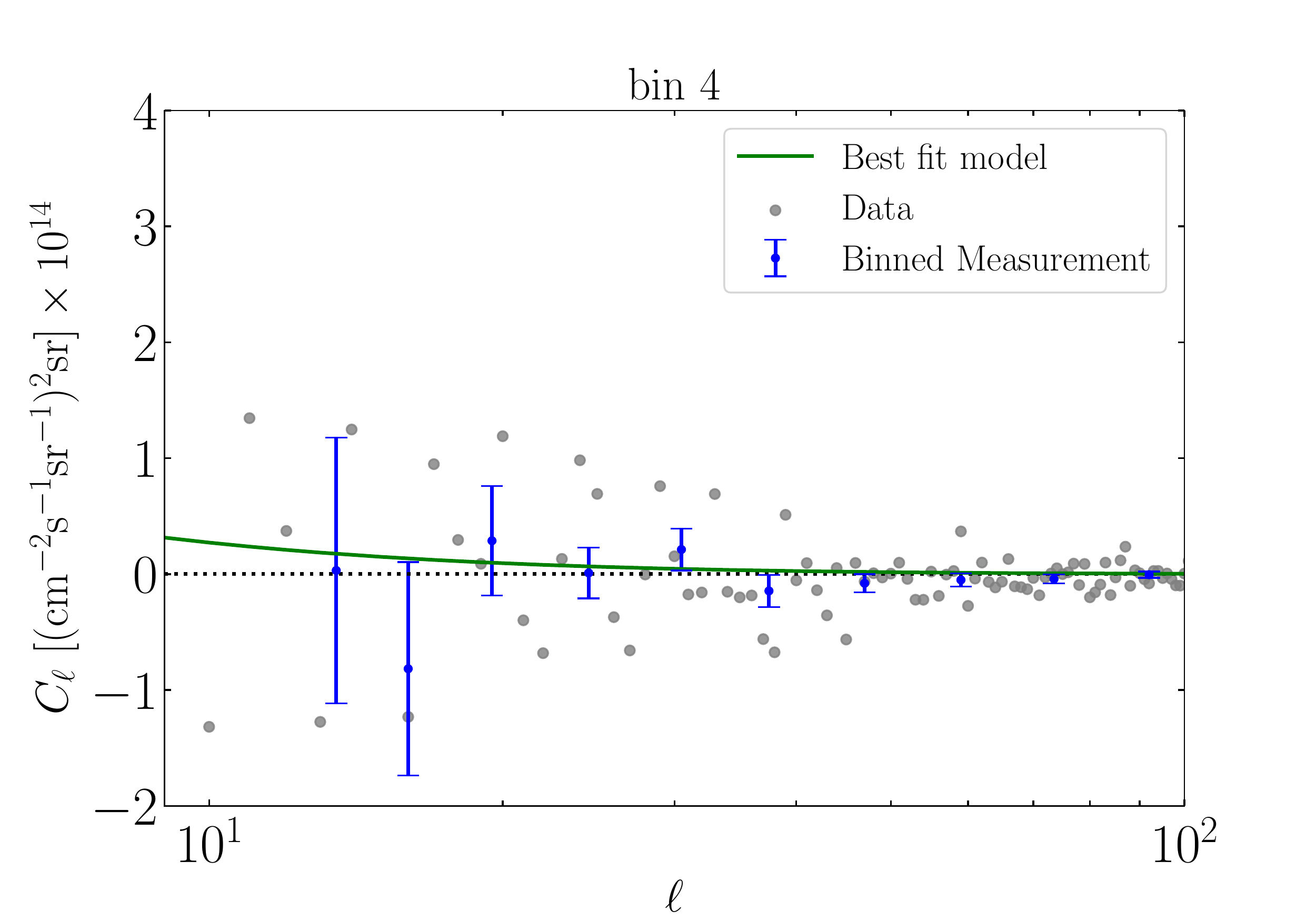}
	\includegraphics[width=4.7cm]{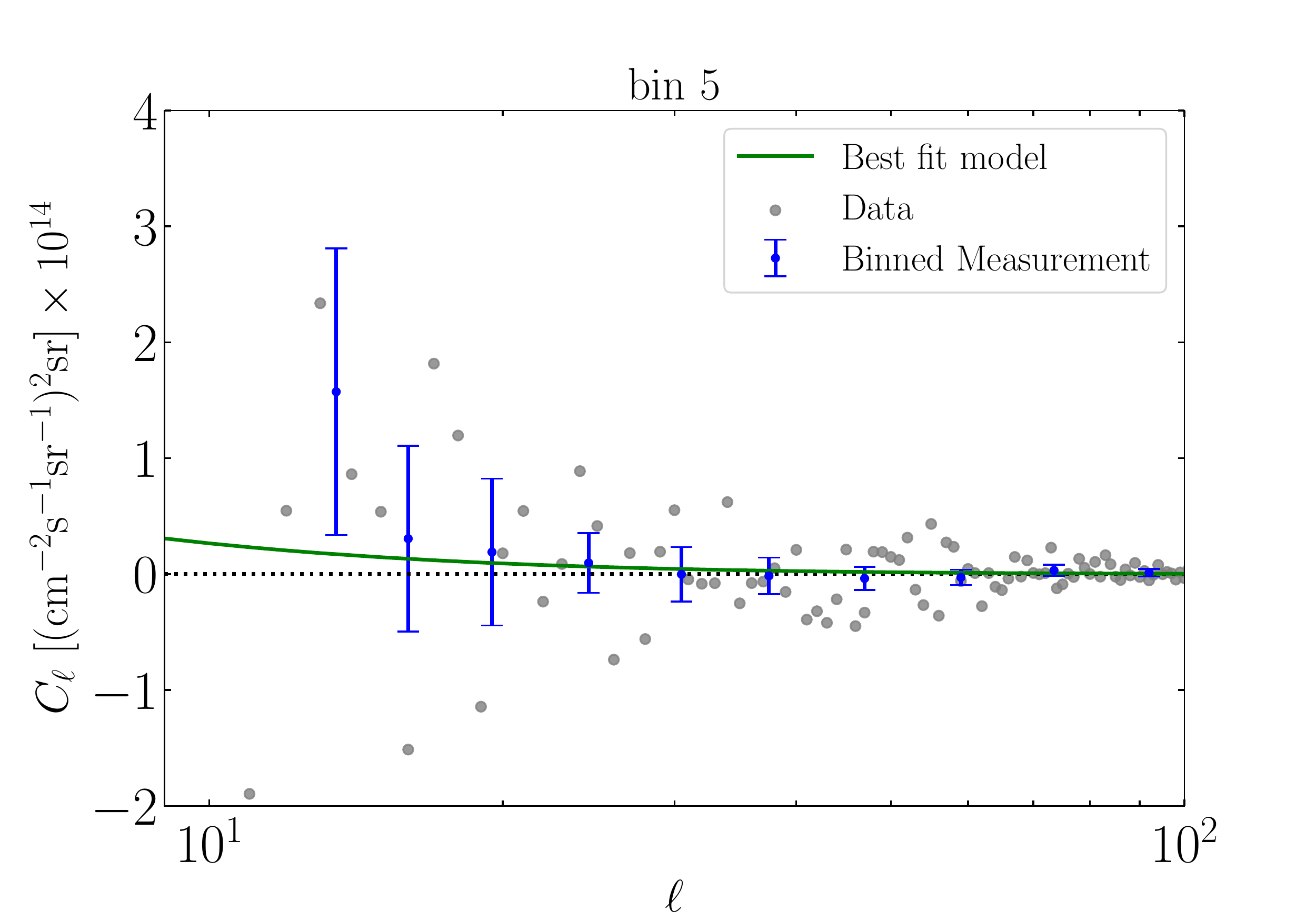}
	\includegraphics[width=4.7cm]{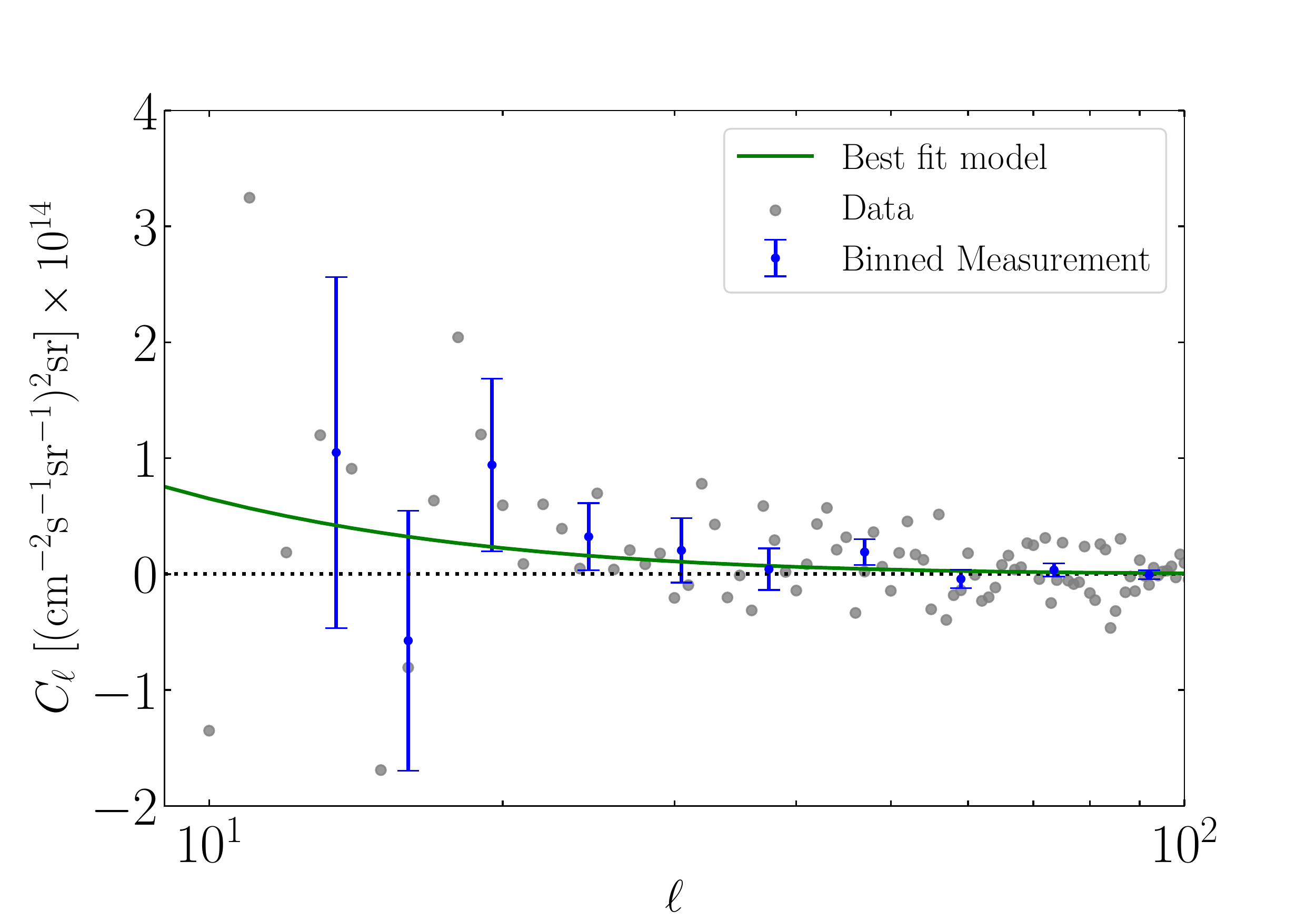}
	\includegraphics[width=4.7cm]{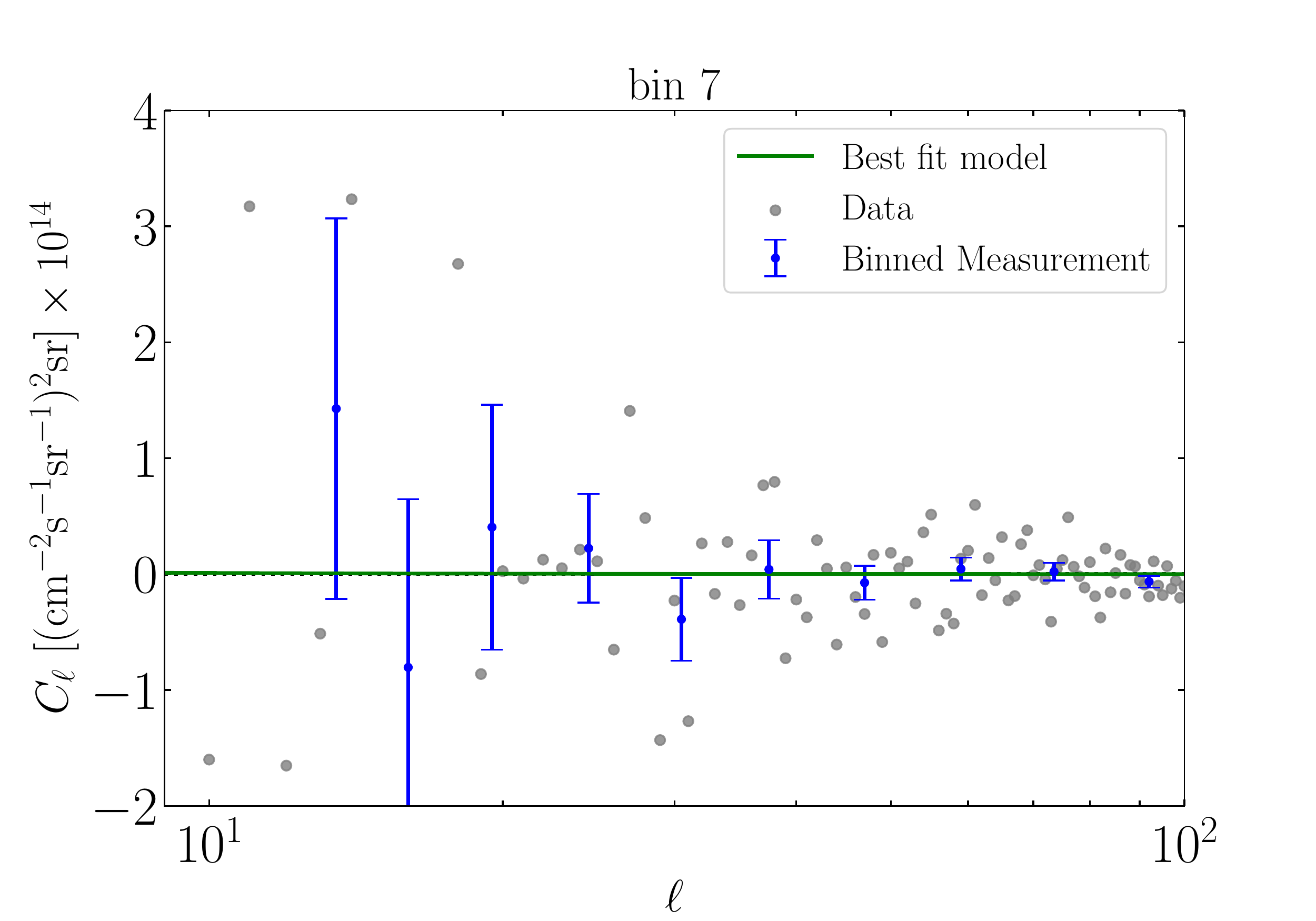}
	\includegraphics[width=4.7cm]{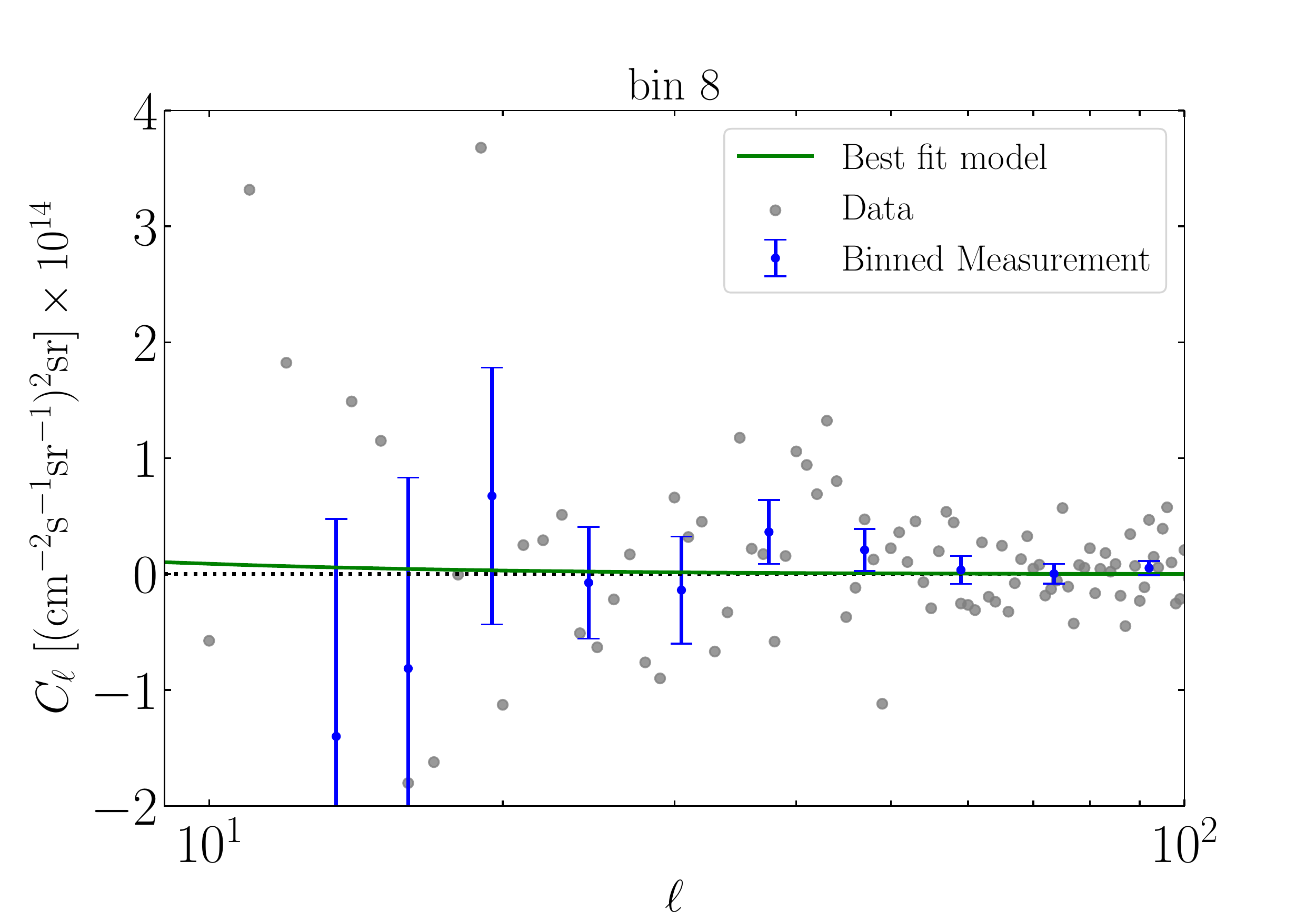}
	\includegraphics[width=4.7cm]{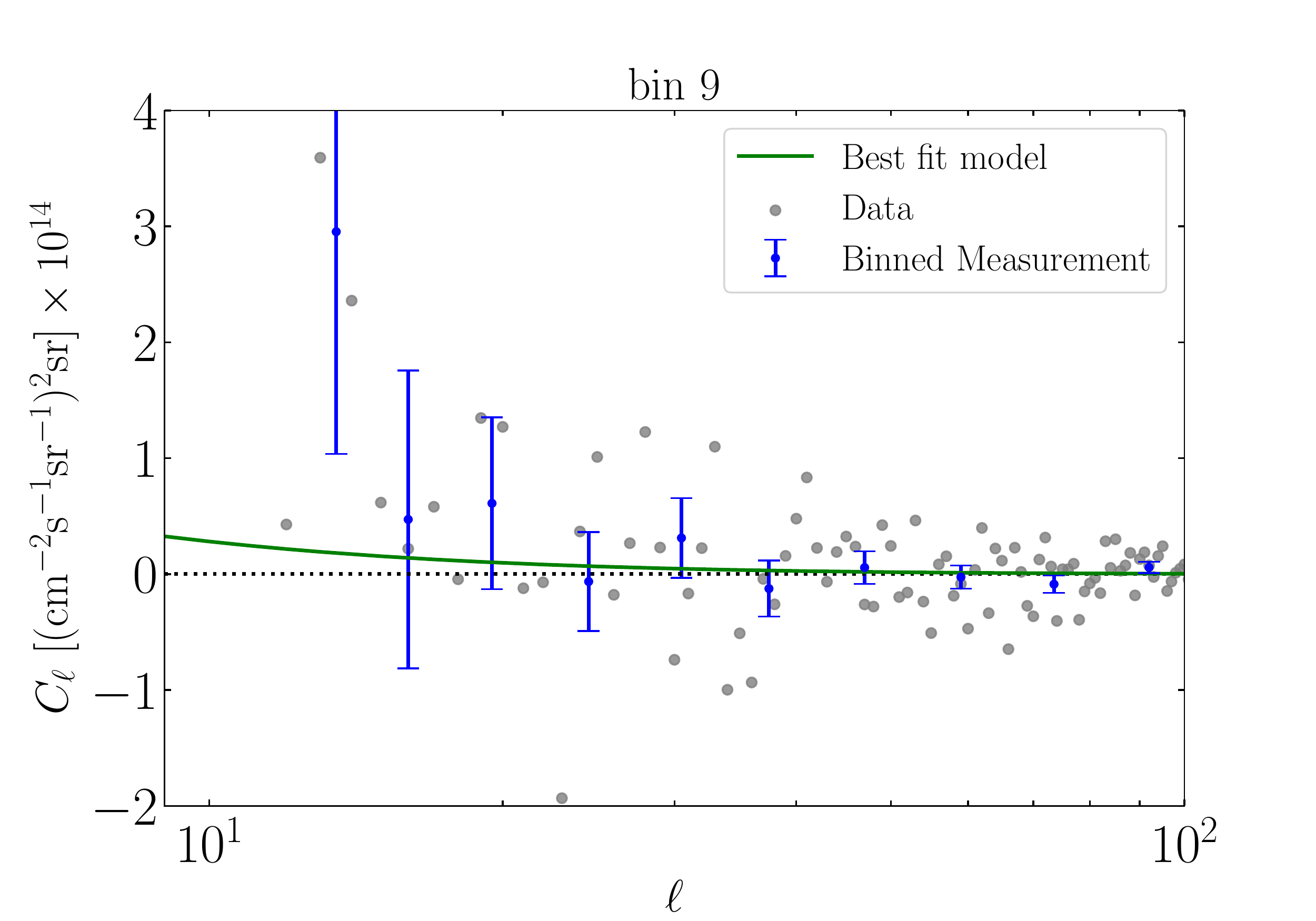}
	\caption{The observed cross-correlation power spectra and the predictions of best-fit models (green solid lines) in each energy bin.
	Gray dots are the original measured power spectra, while blue data points with error bars refer to binned measurements.}
	\label{fig:fitting}
\end{figure*}

\section{Numerical Results}
\label{sec:statistic}

In this section we compute the cross-correlation power spectrum between {\planck} and {\Fermi} \g-ray maps and compare the results with model predictions to quantify the significance of the ISW effect.
We perform a fitting analysis and assume a flat {\lcdm} model with cosmological parameters for the theoretical computation: $\Omega_b h^2 = 0.0224,~\Omega_c h^2 = 0.1201, ~100\theta_{MC} = 1.0409,~\tau = 0.0543,~n_s = 0.9661,~{\rm{ln}}(10^{10}A_s) = 3.0448$, in accordance with the most recent Planck results \citep{Aghanim:2018eyx}.

Referring to \citet{Xia:2015wka}, we use the  PolSpice\footnote{http://www2.iap.fr/users/hivon/software/PolSpice/} statistical tool kit \citep{Szapudi:2001qj,Chon:2003gx,Efstathiou:2003tv,Challinor:2004pr} to estimate the angular power spectrum, which automatically corrects for the effect of the mask.
The accuracy of the PolSpice estimator has been assessed in \citet{Xia:2015wka} by comparing the measured cross-correlation function with the one computed using the popular Landy–Szalay method \citep{Landy:1993yu}.
They were found to be in very good agreement.
PolSpice also provides the covariance matrix for the angular power spectrum.
In Fig. \ref{fig:fitting} we show the measured cross-correlation power spectra between {\planck} and {\Fermi} maps in each energy bin.
Besides these gray observed data points, provided by the PolSpice, we also show the binned measurements with blue points for illustrated purpose.
The error bars come from the diagonal terms of the covariance matrix of each multipole bin.
We have corrected the angular power spectrum by the beam window functions which are associated with PSF effects we mentioned in section \ref{sec:data}.
It is computed from Fermi Tools \texttt{gtpsf} and varies with energy and specific IRF, as the same consideration with \citet{Ammazzalorso:2018evf}.
However, since we only discuss the scales $\ell<512$ and the main ISW information comes from even smaller $\ell$, we found these effects are negligible(see more information in section \ref{sec:statistic}).

\begin{table}
	\caption{Results of the significance of the ISW effect for different data combinations. {$\chi_0^2$ and $\chi_{\min}^2$ are the $\chi^2$ for no ISW effect hypothesis and the minimum for equation (\ref{equ:chi2}). Amplitude is the constraint result of $A$ and $\sigma_{A}$, $\rm SNR$ is signal-to-noise and $\sqrt{\rm TS}$ gives the significant for different bins.}}
	\begin{tabular}{cccccc}
		\hline
		\hline
		Data &Amplitude & ${\rm SNR}/\sigma$ & $\chi_0^2$ & $\chi_{\min}^2$& $\sqrt{\rm TS}/\sigma$ \\
		\hline
		bin1 & $0.28\pm1.49$ &0.2 &210.184 &210.180 & 0.1 \\
		bin2 & $1.99\pm1.09$ &1.8 &182.88 &180.04 &1.8 \\
		bin3 & $0.74\pm1.10$ &0.7 &263.76 &263.30 &0.7 \\
		bin4 & $1.01\pm1.22$ &0.9 &293.01 &292.32 & 0.9\\
		bin5 & $0.72\pm1.44$ &0.5 &272.57 &272.11 &0.7 \\
		bin6 & $2.36\pm1.22$ &1.9 &269.11 &265.76 &1.8 \\
		bin7 & $0.49\pm1.71$ &0.3 &276.08 &275.97 &0.3 \\
		bin8 & $0.78\pm1.84$ &0.4 &300.48 &300.15 &0.6 \\
		bin9 & $1.34\pm1.59$ &0.8 &330.01 &328.79 &1.1 \\
		\hline
		all (binning) & $0.95\pm0.53$ &1.8 &2398.07 &2395.16 &1.7\\
		all (unbinning) & $0.64\pm1.83$ &0.4 &202.42 &202.24 &0.4 \\
		\hline
	\end{tabular}
	\centering
	\label{tab:fitresult}
\end{table}

Firstly, we use the measured cross-correlation power spectra between {\planck} and {\Fermi} maps in each energy bin to constrain the amplitude parameter, through the gaussian likelihood function:
\be
\mathcal{L} = (2\pi)^{-N/2}[{\rm det}(\Gamma_{\ell \ell^\prime})]^{-1/2}{\rm exp}[-\chi^2/2]~,
\ee
where $\Gamma_{\ell \ell \prime}$ is the covariance matrix obtained from the PolSpice estimator. The $\chi^2$ function for each bin is
\be
    \chi^2(A_i)=\left[\hat C_{\ell,i} - A_i C_\ell\right]^T \Gamma_{\ell \ell^\prime}^{-1} \left[\hat C_{\ell^\prime,i} - A_i C_\ell^\prime \right]~,
\label{equ:chi2}
\ee
where $A_i$ is the amplitude parameters in each energy bin.

{The multipole range of cross-correlation power spectrum we used is $\ell=11-512$;
the reason why $\ell$ start with 11 is because the ISW effect is mainly from the large scales but it can be affected by large-scale effects due to an imperfect galactic foreground removal of {\Fermi} maps \cite{Ammazzalorso:2018evf}.
Meanwhile, the Limber approximation we adopted to calculate the cross power spectrum may lead to the uncertainly at the level of 10$\%$ in the small $\ell$.
While large $\ell$ which will not contribute any further improvement in our analysis, also discard since the cross angular power spectrum may be affected by an imperfect PSF correction, especially when the beam window function starts deviating significantly.
So the upper limit on $\ell$ is defined by the condition that the beam window function does not drop below a threshold corresponding approximately to the 68$\%$ containment of the PSF in the specific energy bin.
}

Finally, to quantify the significance of the measurement, we also use another statistic the quantity
\be
\rm{TS} =\chi^2_0-\chi^2_{\min},~
\ee
where $\chi^2_{\min}$ is the minimum $\chi^2$ that calculated in equation (\ref{equ:chi2}), and $\chi^2_0$ is the $\chi^2$ of no ISW effect hypothesis, i.e. of the case $A_i=0$.
TS has an important property to examine the $\chi^2$ distribution with a number of degrees of freedom under the null hypothesis.
By the specific qualities of a descriptive statistic, we can derive the significance level of a measurement based on the measured TS.
In this case, we have only one free parameter here, the significance in sigma is just given by $\sqrt{\rm TS}$.
In table \ref{tab:fitresult} we list the $\sqrt{\rm TS}$ results from different data combinations, another column with the results of signal-to-noise ratio (SNR), which is ${\rm S/N=A/\sigma_{A}}$, also shows in the table.
From the value of every single bins, we can see they are consistent with each other. From now on, we will use the values of SNR to represent the significance level.

Among these nine energy bins, we find that the second and sixth bins provide a significance level around $1.8\sigma$ and $1.9\sigma$ significance of the ISW effect detection, while in other seven energy bins, the measured cross-correlation power spectra show no signal comparing with null detection.
If we look closer, shown in the second panel of Fig. \ref{fig:fitting}, the data points in $30<\ell<40$ seems the main contribution to the $1.8\sigma$ ISW detection.
When we neglect the low multipole information ($\ell < 21$ and $\ell < 31$) in the analysis, we still have about $1.7\sigma$ and $1.4\sigma$ significance of the ISW effect, respectively.
However, the situation will be different in the sixth energy bin.
When we also neglect the large scales data, the significance of ISW effect will quickly drop down to $1.2\sigma$ and $0.4\sigma$, respectively, which means the main signal comes from the larger scales.

Since in this work we compute the cross-correlation signal in nine energy bins, we can use a single power law model, which includes an explicit energy dependence, to normalize the amplitudes of energy bins:
\be
\bar{C}^i_{\ell} = C^i_{\ell} \frac{(E^i_{\rm min} E^i_{\rm max})^{-\alpha_0}} {\Delta E^i}~,
\label{equ:Efit}
\ee
where $E^i_{\rm min}$, $E^i_{\rm max}$ are the minimal and maximal energy in each bin which are listed in table \ref{tab:bins}, $\Delta E = E_{\rm max} - E_{\rm min}$ is the width of the energy bin considered in the cross-correlation analysis, and $\alpha$ is the slope.
\citet{Ackermann:2018wlo} used the angular correlation power spectrum of {\Fermi}-LAT UGRB map to constrain the slope of the single power law model and obtained the tight limit $\alpha=0.13\pm0.03$ ($68\%$ C.L.), which corresponds to $\alpha_0$ in equation (\ref{equ:Efit}): $\alpha = \alpha_0+1$.
Here, we use the same {\Fermi}-LAT UGRB data with slightly different energy separation, which should not significantly change the constraint.
We also try to use our ISW measurements in nine energy bins to constrain this slope, due to the strong degeneracy between the slope and the amplitude and the limited accuracy of data points, we can not obtain any reasonable constraint from the current ISW measurements.
Therefore, here we directly fix the slope to be the best fit value $\alpha=1.13$ and only focus on the constraint of the amplitude.

By combining ISW measurements of the nine energy bins, we can constrain the amplitude parameter:
\be
A_{\rm amp} = 0.95\pm0.53~~({\rm 68\%~C.L.})~,
\ee
This $1.8\sigma$ significance is slightly smaller than those in the single second or sixth energy bin.
After some careful checks, we find that the reason comes from the single power law model.
In this model, we normalize all the amplitudes of energy bins to one single parameter.
Since most of energy bins give non-detection result, the normalized amplitude parameter will be clearly suppressed, which leads that the theoretical predictions in second and sixth energy bin are underestimated.
In the meanwhile, the error bars of data points in these two bins remain the same.
The final significance of ISW detection becomes small and the contributions coming from the second and sixth bin are suppressed, dropping down to $\sqrt{\rm TS}\sim1.3\sigma$.
If only use data points of the second and sixth bins to measure the amplitude parameter, we can obtain the $2.7\sigma$ detection of ISW effect: $A_{\rm amp} = 2.59 \pm 0.95$ at 68\% confidence level.

Furthermore, we combine all the UGRB information into one single energy bin, which is basically one map of full energy band, to measure the cross-correlation power spectrum between {\planck} and {\Fermi}-LAT UGRB maps.
Different from the binning results above, using one energy bin will significantly lose the LSS information and only give very weak significance, about $0.3\sigma$ confidence level.
\begin{figure*}
	\centering
	\includegraphics[width=6.5cm]{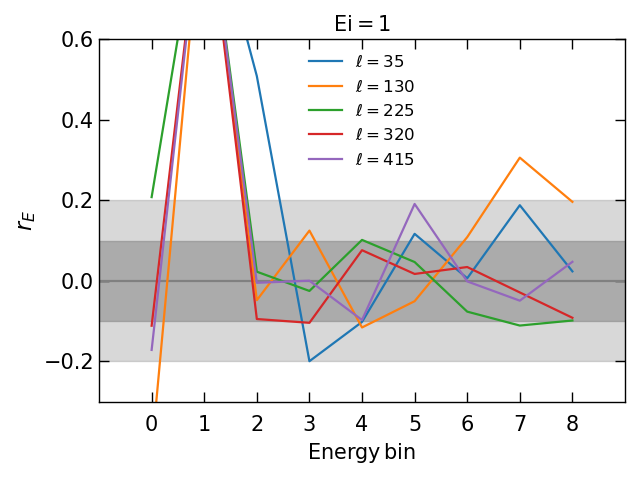}
	\includegraphics[width=6.5cm]{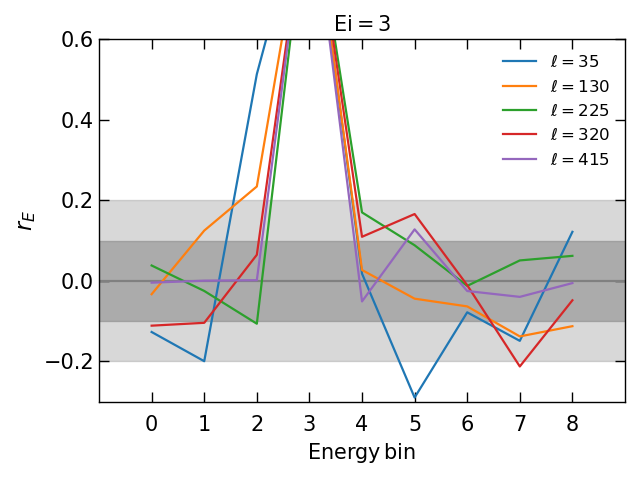}
	\includegraphics[width=6.5cm]{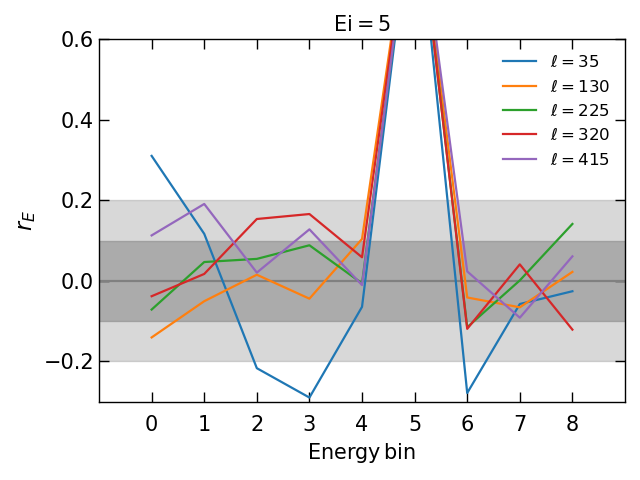}
	\includegraphics[width=6.5cm]{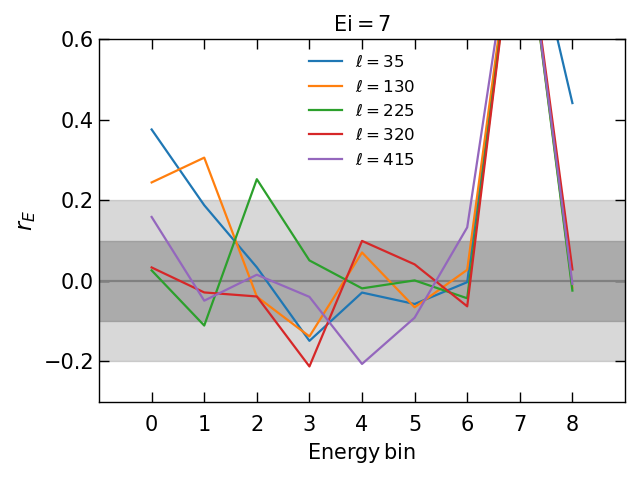}
	\caption{Four energy slices of cross-correlation coefficient as defined by equation \ref{equ:re}.
	The title of $E_i$ refers to $i$ or $j$ energy bin of $r_E^{i,j}$, the other one represents on the horizontal scale of each panel.
	Each coloured line stands different multipole in our analysis range.
	The peaks are diagonal covariance that namely, $r_E^{i,i}$, which equals to 1 by definition.
	Therefore other points are off-diagonal terms of the covariance matrix.}
	\label{fig:re}
\end{figure*}

{
For the sake of validation by summing energy bins together when estimate all bins $\chi^2$ in Tab. \ref{tab:fitresult}, we implement a correlation coefficient to describe the covariance between different energy bins.
This Gaussian estimator can be determined by
\begin{equation}
    r_E^{i,j}=\frac{\Gamma_{\ell \ell}^{T, \gamma_i \gamma_j}}{\sqrt{\Gamma_{\ell \ell}^{T, \gamma_i \gamma_i}\Gamma_{\ell \ell}^{T, \gamma_j \gamma_j}}},
    \label{equ:re}
\end{equation}
which the covariance at fixed multipole above in one energy bin is calculating by
\begin{equation}
    \Gamma_{\ell \ell}^{T, \gamma_i \gamma_j} \equiv \frac{C_\ell^{TT}C_\ell^{\gamma_i \gamma_j}+C_\ell^{T\gamma_i}C_\ell^{T\gamma_j}}{(2\ell+1) f_{sky}}.
    \label{equ:cov}
\end{equation}
The $f_{sky}$ is the coverage percentage of CMB map over full sky map.\\
With fixed multipole, the $r_E$ illustrates the off-diagonal elements of the covariance between \g-ray maps and CMB map.
There are four arbitrary energy slices among 9 bins of cross-correlation coefficient $r_E$ in Fig. \ref{fig:re}.
For instance, the left-top one is showing the result of first bin with the other 9 bins.
Those peaks are the diagonal of covariance which equal to 1 by definition. Other points are off-diagonal terms.
They show the off-diagonal elements of the covariance matrix in high multipole are mainly below 10\% deviation, which means the independence between them.
Despite the correlation are higher at low $\ell$ than other place, which means the different energy bin maps contribute similar way to ISW signal, it allows us to analyze the combining detection by summing them up because of most of them are under 10\% deviation.}

\section{Systematic Tests}
\label{sec:null}

To check the robustness of the results, we performed further tests using mock catalogs with no cross correlation with CMB temperature fluctuations, verifying that the computed cross-correlation power spectra are compatible with a null signal.

In each energy bin, we create a Monte Carlo catalog that redistributed the galaxies of the catalog randomly over the sky area and remains the same total flux with the original map.
In this case the new catalog contains no intrinsic clustering.
Following the procedures above, we could use these random mock catalog to cross-correlate with the {\planck} map and measure the normalized amplitude parameter.
As we expected, the obtained significance is very close to zero, ${\rm S/N} < 0.1\sigma$. The clustering feature disappears when using the random mock catalog.

\begin{table}
	\caption{Results of significance of ISW detection using different galactic cuts.}
	\begin{tabular}{cccc}
		\hline
		\hline
		Samples & Amplitude & SNR & $\sqrt{\rm TS}$\\
		\hline
		$|b|<20^{\circ}$ & $0.49 \pm 0.51$ & 1.0 & 1.0\\
		$|b|<25^{\circ}$ & $0.81 \pm 0.49$ & 1.7 & 1.6\\
		$|b|<30^{\circ}$ & $0.95 \pm 0.53$ & 1.8 & 1.8\\
		$|b|<35^{\circ}$ & $0.63 \pm 0.65$ & 1.0 & 0.9\\
		$|b|<40^{\circ}$ & $0.34 \pm 0.79$ & 0.4 & 0.4\\
		\hline
	\end{tabular}
	\centering
	\label{tab:galacticcut}
\end{table}

Furthermore, we vary the galactic cut from $20^{\circ}$ to $40^{\circ}$ in order to test the robustness of our results with the galactic mask.
Then we computed the residuals of real {\Fermi} maps using different galactic cut on the mask in all energy bins and measured the cross-correlation power spectra with the {\planck} map to estimate the significance of ISW detection.
In table \ref{tab:galacticcut}, we list the obtained significance using different galactic cuts on the mask.
As we expected, the highest significance comes from the {\Fermi} map with the galactic cut $b=30^{\circ}$, which is consistent with our previous result \citep{Xia:2011ax}.
When we increase the galactic cut area on the mask, the UGRB information will be significantly lost, for example, the area of {\Fermi} map with the $|b|<40^{\circ}$ cut is only half of that of the map with $|b|<20^{\circ}$ cut.
Therefore, the obtained error bar of the normalized amplitude parameter becomes very large.
Consequently, the values of SNR are smaller.
When decreasing the galactic cut area, the galactic contamination will be strong and the systematic errors could be larger, which caused the significance lower than before.

\section{Conclusions}
\label{sec:concl}

In this paper, we use the UGRB data in the energy range [0.631, 1000]GeV from the latest {\Fermi}-LAT \g-ray observation to estimate the cross-correlation power spectra with the {\planck} CMB map, and investigate its capability on the detection of the well-known ISW effect.
Following the procedure of foreground cleaning in \citet{Xia:2015wka,Cuoco:2017bpv,Ackermann:2018wlo}, for the first time, we obtain a positive evidence with about $1.8\sigma$ significance for the ISW detection from the UGRB information of the {\Fermi} P8 data.

Here we summarize our main conclusions in more detail:

\begin{itemize}
    \item We report the positive evidence at about $1.8\sigma$ confidence level from the cross-correlation power spectra of {\Fermi} \g-ray maps and {\planck} CMB map in the energy range [1.202, 2.290] GeV and [17.38, 36.31] GeV.
    When combining these two energy bins together, we obtain a $2.7\sigma$ detection of the ISW effect: $A_{\rm amp}=2.59\pm0.95$ (68\% C.L.).

    \item Then we use the single power-law model to normalize the amplitude in each energy bin [equation (\ref{equ:Efit})] and estimate the normalized amplitude parameter from all nine energy bins together.
    For the first time, we obtain the significance of the ISW effect at about $1.8\sigma$ confidence level: $A_{\rm amp} = 0.95\pm0.53$. When comparing with the result of \citet{Xia:2011ax}, the main improvement comes for the updated {\Fermi}-LAT data.

    \item Finally, we implement a null hypothesis to test the cross-correlation between CMB and the UGRB of {\Fermi} by randomizing the {\Fermi} maps of nine energy bins, and obtain that these randomized mock {\Fermi} maps contain no intrinsic clustering.
    This null test confirm the above signal with its near zero significant.
    The correlation coefficient as an indicator to expose the relation between energy bins, help us to confirm the calculation of combination signal.

    \item Furthermore, we generate several mask maps with different galactic cut and check the influence of these galactic cut on the obtained significance of ISW effect.
    Similar with our previous work, we find the galactic cut $|b|>30^{\circ}$ could give the highest signal to noise ratio.
\end{itemize}

\section*{Acknowledgements}

We would like to thank Marco Regis for helpful discussions.
This work was supported by the National Science Foundation of China under grants Nos. U1931202, 11633001, and 11690023 and the National Key R\&D Program of
China No. 2017YFA0402600.

\bibliographystyle{elsarticle}
\bibliography{references}

\begin{thebibliography}{45}
\expandafter\ifx\csname natexlab\endcsname\relax\def\natexlab#1{#1}\fi
\providecommand{\url}[1]{\texttt{#1}}
\providecommand{\href}[2]{#2}
\providecommand{\path}[1]{#1}
\providecommand{\DOIprefix}{doi:}
\providecommand{\ArXivprefix}{arXiv:}
\providecommand{\URLprefix}{URL: }
\providecommand{\Pubmedprefix}{pmid:}
\providecommand{\doi}[1]{\href{http://dx.doi.org/#1}{\path{#1}}}
\providecommand{\Pubmed}[1]{\href{pmid:#1}{\path{#1}}}
\providecommand{\bibinfo}[2]{#2}
\ifx\xfnm\relax \def\xfnm[#1]{\unskip,\space#1}\fi
\bibitem[{Aghanim et~al.(2018)}]{Aghanim:2018eyx}
\bibinfo{author}{N.~Aghanim}, et~al. (\bibinfo{collaboration}{Planck}),
\newblock \bibinfo{title}{{Planck 2018 results. VI. Cosmological parameters}}
  (\bibinfo{year}{2018}). \href{http://arxiv.org/abs/1807.06209}{{\tt
  arXiv:1807.06209}}.
\bibitem[{Eisenstein et~al.(2005)}]{Eisenstein:2005su}
\bibinfo{author}{D.~J. Eisenstein}, et~al. (\bibinfo{collaboration}{SDSS}),
\newblock \bibinfo{title}{{Detection of the Baryon Acoustic Peak in the
  Large-Scale Correlation Function of SDSS Luminous Red Galaxies}},
\newblock \bibinfo{journal}{Astrophys. J.} \bibinfo{volume}{633}
  (\bibinfo{year}{2005}) \bibinfo{pages}{560--574}.
  \DOIprefix\doi{10.1086/466512}.
  \href{http://arxiv.org/abs/astro-ph/0501171}{{\tt arXiv:astro-ph/0501171}}.
\bibitem[{Riess et~al.(2004)}]{Riess:2004nr}
\bibinfo{author}{A.~G. Riess}, et~al. (\bibinfo{collaboration}{Supernova Search
  Team}),
\newblock \bibinfo{title}{{Type Ia supernova discoveries at z > 1 from the
  Hubble Space Telescope: Evidence for past deceleration and constraints on
  dark energy evolution}},
\newblock \bibinfo{journal}{Astrophys. J.} \bibinfo{volume}{607}
  (\bibinfo{year}{2004}) \bibinfo{pages}{665--687}.
  \DOIprefix\doi{10.1086/383612}.
  \href{http://arxiv.org/abs/astro-ph/0402512}{{\tt arXiv:astro-ph/0402512}}.
\bibitem[{Bacon et~al.(2000)Bacon, Refregier, and Ellis}]{Bacon:2000sy}
\bibinfo{author}{D.~J. Bacon}, \bibinfo{author}{A.~R. Refregier},
  \bibinfo{author}{R.~S. Ellis},
\newblock \bibinfo{title}{{Detection of weak gravitational lensing by
  large-scale structure}},
\newblock \bibinfo{journal}{Mon. Not. Roy. Astron. Soc.} \bibinfo{volume}{318}
  (\bibinfo{year}{2000}) \bibinfo{pages}{625}.
  \DOIprefix\doi{10.1046/j.1365-8711.2000.03851.x}.
  \href{http://arxiv.org/abs/astro-ph/0003008}{{\tt arXiv:astro-ph/0003008}}.
\bibitem[{Sachs and Wolfe(1967)}]{Sachs:1967er}
\bibinfo{author}{R.~K. Sachs}, \bibinfo{author}{A.~M. Wolfe},
\newblock \bibinfo{title}{{Perturbations of a cosmological model and angular
  variations of the microwave background}},
\newblock \bibinfo{journal}{Astrophys. J.} \bibinfo{volume}{147}
  (\bibinfo{year}{1967}) \bibinfo{pages}{73--90}.
  \DOIprefix\doi{10.1007/s10714-007-0448-9}, \bibinfo{note}{[Gen. Rel.
  Grav.39,1929(2007)]}.
\bibitem[{Crittenden and Turok(1996)}]{Crittenden:1995ak}
\bibinfo{author}{R.~G. Crittenden}, \bibinfo{author}{N.~Turok},
\newblock \bibinfo{title}{{Looking for Lambda with the Rees-Sciama effect}},
\newblock \bibinfo{journal}{Phys. Rev. Lett.} \bibinfo{volume}{76}
  (\bibinfo{year}{1996}) \bibinfo{pages}{575}.
  \DOIprefix\doi{10.1103/PhysRevLett.76.575}.
  \href{http://arxiv.org/abs/astro-ph/9510072}{{\tt arXiv:astro-ph/9510072}}.
\bibitem[{Boughn et~al.(1998)Boughn, Crittenden, and Turok}]{Boughn:1997vs}
\bibinfo{author}{S.~P. Boughn}, \bibinfo{author}{R.~G. Crittenden},
  \bibinfo{author}{N.~G. Turok},
\newblock \bibinfo{title}{{Correlations between the cosmic X-ray and microwave
  backgrounds: Constraints on a cosmological constant}},
\newblock \bibinfo{journal}{New Astron.} \bibinfo{volume}{3}
  (\bibinfo{year}{1998}) \bibinfo{pages}{275--291}.
  \DOIprefix\doi{10.1016/S1384-1076(98)00009-8}.
  \href{http://arxiv.org/abs/astro-ph/9704043}{{\tt arXiv:astro-ph/9704043}}.
\bibitem[{Afshordi et~al.(2004)Afshordi, Loh, and Strauss}]{Afshordi:2003xu}
\bibinfo{author}{N.~Afshordi}, \bibinfo{author}{Y.-S. Loh},
  \bibinfo{author}{M.~A. Strauss},
\newblock \bibinfo{title}{{Cross - correlation of the Cosmic Microwave
  Background with the 2MASS galaxy survey: Signatures of dark energy, hot gas,
  and point sources}},
\newblock \bibinfo{journal}{Phys. Rev.} \bibinfo{volume}{D69}
  (\bibinfo{year}{2004}) \bibinfo{pages}{083524}.
  \DOIprefix\doi{10.1103/PhysRevD.69.083524}.
  \href{http://arxiv.org/abs/astro-ph/0308260}{{\tt arXiv:astro-ph/0308260}}.
\bibitem[{Francis and Peacock(2010)}]{Francis:2009pt}
\bibinfo{author}{C.~L. Francis}, \bibinfo{author}{J.~A. Peacock},
\newblock \bibinfo{title}{{An estimate of the local ISW signal, and its impact
  on CMB anomalies}},
\newblock \bibinfo{journal}{Mon. Not. Roy. Astron. Soc.} \bibinfo{volume}{406}
  (\bibinfo{year}{2010}) \bibinfo{pages}{14}.
  \DOIprefix\doi{10.1111/j.1365-2966.2010.16866.x}.
  \href{http://arxiv.org/abs/0909.2495}{{\tt arXiv:0909.2495}}.
\bibitem[{Dupe et~al.(2011)Dupe, Rassat, Starck, and Fadili}]{Dupe:2010zs}
\bibinfo{author}{F.~X. Dupe}, \bibinfo{author}{A.~Rassat},
  \bibinfo{author}{J.~L. Starck}, \bibinfo{author}{M.~J. Fadili},
\newblock \bibinfo{title}{{Measuring the Integrated Sachs-Wolfe Effect}},
\newblock \bibinfo{journal}{Astron. Astrophys.} \bibinfo{volume}{534}
  (\bibinfo{year}{2011}) \bibinfo{pages}{A51}.
  \DOIprefix\doi{10.1051/0004-6361/201015893}.
  \href{http://arxiv.org/abs/1010.2192}{{\tt arXiv:1010.2192}}.
\bibitem[{Nolta et~al.(2004)}]{Nolta:2003uy}
\bibinfo{author}{M.~R. Nolta}, et~al. (\bibinfo{collaboration}{WMAP}),
\newblock \bibinfo{title}{{First year Wilkinson Microwave Anisotropy Probe
  (WMAP) observations: Dark energy induced correlation with radio sources}},
\newblock \bibinfo{journal}{Astrophys. J.} \bibinfo{volume}{608}
  (\bibinfo{year}{2004}) \bibinfo{pages}{10--15}.
  \DOIprefix\doi{10.1086/386536}.
  \href{http://arxiv.org/abs/astro-ph/0305097}{{\tt arXiv:astro-ph/0305097}}.
\bibitem[{Scranton et~al.(2003)}]{Scranton:2003in}
\bibinfo{author}{R.~Scranton}, et~al. (\bibinfo{collaboration}{SDSS}),
\newblock \bibinfo{title}{{Physical evidence for dark energy}}
  (\bibinfo{year}{2003}). \href{http://arxiv.org/abs/astro-ph/0307335}{{\tt
  arXiv:astro-ph/0307335}}.
\bibitem[{Fosalba et~al.(2003)Fosalba, Gaztanaga, and
  Castander}]{Fosalba:2003ge}
\bibinfo{author}{P.~Fosalba}, \bibinfo{author}{E.~Gaztanaga},
  \bibinfo{author}{F.~Castander},
\newblock \bibinfo{title}{{Detection of the ISW and SZ effects from the
  CMB-galaxy correlation}},
\newblock \bibinfo{journal}{Astrophys. J.} \bibinfo{volume}{597}
  (\bibinfo{year}{2003}) \bibinfo{pages}{L89--92}.
  \DOIprefix\doi{10.1086/379848}.
  \href{http://arxiv.org/abs/astro-ph/0307249}{{\tt arXiv:astro-ph/0307249}}.
\bibitem[{Granett et~al.(2008)Granett, Neyrinck, and Szapudi}]{Granett:2008ju}
\bibinfo{author}{B.~R. Granett}, \bibinfo{author}{M.~C. Neyrinck},
  \bibinfo{author}{I.~Szapudi},
\newblock \bibinfo{title}{{An Imprint of Super-Structures on the Microwave
  Background due to the Integrated Sachs-Wolfe Effect}},
\newblock \bibinfo{journal}{Astrophys. J.} \bibinfo{volume}{683}
  (\bibinfo{year}{2008}) \bibinfo{pages}{L99--L102}.
  \DOIprefix\doi{10.1086/591670}. \href{http://arxiv.org/abs/0805.3695}{{\tt
  arXiv:0805.3695}}.
\bibitem[{Chen and Schwarz(2016)}]{Chen:2015wga}
\bibinfo{author}{S.~Chen}, \bibinfo{author}{D.~J. Schwarz},
\newblock \bibinfo{title}{{Angular two-point correlation of NVSS galaxies
  revisited}},
\newblock \bibinfo{journal}{Astron. Astrophys.} \bibinfo{volume}{591}
  (\bibinfo{year}{2016}) \bibinfo{pages}{A135}.
  \DOIprefix\doi{10.1051/0004-6361/201526956}.
  \href{http://arxiv.org/abs/1507.02160}{{\tt arXiv:1507.02160}}.
\bibitem[{Goto et~al.(2012)Goto, Szapudi, and Granett}]{Goto:2012yc}
\bibinfo{author}{T.~Goto}, \bibinfo{author}{I.~Szapudi}, \bibinfo{author}{B.~R.
  Granett},
\newblock \bibinfo{title}{{Cross-correlation of WISE Galaxies with the Cosmic
  Microwave Background}},
\newblock \bibinfo{journal}{Mon. Not. Roy. Astron. Soc.} \bibinfo{volume}{422}
  (\bibinfo{year}{2012}) \bibinfo{pages}{L77--L81}.
  \DOIprefix\doi{10.1111/j.1745-3933.2012.01240.x}.
  \href{http://arxiv.org/abs/1202.5306}{{\tt arXiv:1202.5306}}.
\bibitem[{Ferraro et~al.(2015)Ferraro, Sherwin, and Spergel}]{Ferraro:2014msa}
\bibinfo{author}{S.~Ferraro}, \bibinfo{author}{B.~D. Sherwin},
  \bibinfo{author}{D.~N. Spergel},
\newblock \bibinfo{title}{{WISE measurement of the integrated Sachs-Wolfe
  effect}},
\newblock \bibinfo{journal}{Phys. Rev.} \bibinfo{volume}{D91}
  (\bibinfo{year}{2015}) \bibinfo{pages}{083533}.
  \DOIprefix\doi{10.1103/PhysRevD.91.083533}.
  \href{http://arxiv.org/abs/1401.1193}{{\tt arXiv:1401.1193}}.
\bibitem[{Boughn et~al.(2002)Boughn, Crittenden, and Koehrsen}]{Boughn:2002bs}
\bibinfo{author}{S.~P. Boughn}, \bibinfo{author}{R.~G. Crittenden},
  \bibinfo{author}{G.~P. Koehrsen},
\newblock \bibinfo{title}{{The Large scale structure of the x-ray background
  and its cosmological implications}},
\newblock \bibinfo{journal}{Astrophys. J.} \bibinfo{volume}{580}
  (\bibinfo{year}{2002}) \bibinfo{pages}{672--684}.
  \DOIprefix\doi{10.1086/343861}.
  \href{http://arxiv.org/abs/astro-ph/0208153}{{\tt arXiv:astro-ph/0208153}}.
\bibitem[{Boughn and Crittenden(2004)}]{Boughn:2003yz}
\bibinfo{author}{S.~Boughn}, \bibinfo{author}{R.~Crittenden},
\newblock \bibinfo{title}{{A Correlation of the cosmic microwave sky with large
  scale structure}},
\newblock \bibinfo{journal}{Nature} \bibinfo{volume}{427}
  (\bibinfo{year}{2004}) \bibinfo{pages}{45--47}.
  \DOIprefix\doi{10.1038/nature02139}.
  \href{http://arxiv.org/abs/astro-ph/0305001}{{\tt arXiv:astro-ph/0305001}}.
\bibitem[{Giannantonio et~al.(2008)Giannantonio, Scranton, Crittenden, Nichol,
  Boughn, Myers, and Richards}]{Giannantonio:2008zi}
\bibinfo{author}{T.~Giannantonio}, \bibinfo{author}{R.~Scranton},
  \bibinfo{author}{R.~G. Crittenden}, \bibinfo{author}{R.~C. Nichol},
  \bibinfo{author}{S.~P. Boughn}, \bibinfo{author}{A.~D. Myers},
  \bibinfo{author}{G.~T. Richards},
\newblock \bibinfo{title}{{Combined analysis of the integrated Sachs-Wolfe
  effect and cosmological implications}},
\newblock \bibinfo{journal}{Phys. Rev.} \bibinfo{volume}{D77}
  (\bibinfo{year}{2008}) \bibinfo{pages}{123520}.
  \DOIprefix\doi{10.1103/PhysRevD.77.123520}.
  \href{http://arxiv.org/abs/0801.4380}{{\tt arXiv:0801.4380}}.
\bibitem[{Xia et~al.(2011)Xia, Cuoco, Branchini, Fornasa, and
  Viel}]{Xia:2011ax}
\bibinfo{author}{J.-Q. Xia}, \bibinfo{author}{A.~Cuoco},
  \bibinfo{author}{E.~Branchini}, \bibinfo{author}{M.~Fornasa},
  \bibinfo{author}{M.~Viel},
\newblock \bibinfo{title}{{A cross-correlation study of the Fermi-LAT
  $\gamma$-ray diffuse extragalactic signal}},
\newblock \bibinfo{journal}{Mon. Not. Roy. Astron. Soc.} \bibinfo{volume}{416}
  (\bibinfo{year}{2011}) \bibinfo{pages}{2247--2264}.
  \DOIprefix\doi{10.1111/j.1365-2966.2011.19200.x}.
  \href{http://arxiv.org/abs/1103.4861}{{\tt arXiv:1103.4861}}.
\bibitem[{Khosravi et~al.(2016)Khosravi, Mollazadeh, and
  Baghram}]{Khosravi:2015boa}
\bibinfo{author}{S.~Khosravi}, \bibinfo{author}{A.~Mollazadeh},
  \bibinfo{author}{S.~Baghram},
\newblock \bibinfo{title}{{ISW-galaxy cross correlation: a probe of dark energy
  clustering and distribution of dark matter tracers}},
\newblock \bibinfo{journal}{JCAP} \bibinfo{volume}{1609} (\bibinfo{year}{2016})
  \bibinfo{pages}{003}. \DOIprefix\doi{10.1088/1475-7516/2016/09/003}.
  \href{http://arxiv.org/abs/1510.01720}{{\tt arXiv:1510.01720}}.
\bibitem[{Ade et~al.(2014)}]{Ade:2013dsi}
\bibinfo{author}{P.~A.~R. Ade}, et~al. (\bibinfo{collaboration}{Planck}),
\newblock \bibinfo{title}{{Planck 2013 results. XIX. The integrated Sachs-Wolfe
  effect}},
\newblock \bibinfo{journal}{Astron. Astrophys.} \bibinfo{volume}{571}
  (\bibinfo{year}{2014}) \bibinfo{pages}{A19}.
  \DOIprefix\doi{10.1051/0004-6361/201321526}.
  \href{http://arxiv.org/abs/1303.5079}{{\tt arXiv:1303.5079}}.
\bibitem[{Ade et~al.(2016)}]{Ade:2015dva}
\bibinfo{author}{P.~A.~R. Ade}, et~al. (\bibinfo{collaboration}{Planck}),
\newblock \bibinfo{title}{{Planck 2015 results. XXI. The integrated Sachs-Wolfe
  effect}},
\newblock \bibinfo{journal}{Astron. Astrophys.} \bibinfo{volume}{594}
  (\bibinfo{year}{2016}) \bibinfo{pages}{A21}.
  \DOIprefix\doi{10.1051/0004-6361/201525831}.
  \href{http://arxiv.org/abs/1502.01595}{{\tt arXiv:1502.01595}}.
\bibitem[{Stölzner et~al.(2018)Stölzner, Cuoco, Lesgourgues, and
  Bilicki}]{Stolzner:2017ged}
\bibinfo{author}{B.~Stölzner}, \bibinfo{author}{A.~Cuoco},
  \bibinfo{author}{J.~Lesgourgues}, \bibinfo{author}{M.~Bilicki},
\newblock \bibinfo{title}{{Updated tomographic analysis of the integrated
  Sachs-Wolfe effect and implications for dark energy}},
\newblock \bibinfo{journal}{Phys. Rev.} \bibinfo{volume}{D97}
  (\bibinfo{year}{2018}) \bibinfo{pages}{063506}.
  \DOIprefix\doi{10.1103/PhysRevD.97.063506}.
  \href{http://arxiv.org/abs/1710.03238}{{\tt arXiv:1710.03238}}.
\bibitem[{Camera et~al.(2015)Camera, Fornasa, Fornengo, and
  Regis}]{Camera:2014rja}
\bibinfo{author}{S.~Camera}, \bibinfo{author}{M.~Fornasa},
  \bibinfo{author}{N.~Fornengo}, \bibinfo{author}{M.~Regis},
\newblock \bibinfo{title}{{Tomographic-spectral approach for dark matter
  detection in the cross-correlation between cosmic shear and diffuse
  $\gamma$-ray emission}},
\newblock \bibinfo{journal}{JCAP} \bibinfo{volume}{1506} (\bibinfo{year}{2015})
  \bibinfo{pages}{029}. \DOIprefix\doi{10.1088/1475-7516/2015/06/029}.
  \href{http://arxiv.org/abs/1411.4651}{{\tt arXiv:1411.4651}}.
\bibitem[{Cuoco et~al.(2015)Cuoco, Xia, Regis, Branchini, Fornengo, and
  Viel}]{Cuoco:2015rfa}
\bibinfo{author}{A.~Cuoco}, \bibinfo{author}{J.-Q. Xia},
  \bibinfo{author}{M.~Regis}, \bibinfo{author}{E.~Branchini},
  \bibinfo{author}{N.~Fornengo}, \bibinfo{author}{M.~Viel},
\newblock \bibinfo{title}{{Dark Matter Searches in the Gamma-ray Extragalactic
  Background via Cross-correlations With Galaxy Catalogs}},
\newblock \bibinfo{journal}{Astrophys. J. Suppl.} \bibinfo{volume}{221}
  (\bibinfo{year}{2015}) \bibinfo{pages}{29}.
  \DOIprefix\doi{10.1088/0067-0049/221/2/29}.
  \href{http://arxiv.org/abs/1506.01030}{{\tt arXiv:1506.01030}}.
\bibitem[{Regis et~al.(2015)Regis, Xia, Cuoco, Branchini, Fornengo, and
  Viel}]{Regis:2015zka}
\bibinfo{author}{M.~Regis}, \bibinfo{author}{J.-Q. Xia},
  \bibinfo{author}{A.~Cuoco}, \bibinfo{author}{E.~Branchini},
  \bibinfo{author}{N.~Fornengo}, \bibinfo{author}{M.~Viel},
\newblock \bibinfo{title}{{Particle dark matter searches outside the Local
  Group}},
\newblock \bibinfo{journal}{Phys. Rev. Lett.} \bibinfo{volume}{114}
  (\bibinfo{year}{2015}) \bibinfo{pages}{241301}.
  \DOIprefix\doi{10.1103/PhysRevLett.114.241301}.
  \href{http://arxiv.org/abs/1503.05922}{{\tt arXiv:1503.05922}}.
\bibitem[{Xia et~al.(2015)Xia, Cuoco, Branchini, and Viel}]{Xia:2015wka}
\bibinfo{author}{J.-Q. Xia}, \bibinfo{author}{A.~Cuoco},
  \bibinfo{author}{E.~Branchini}, \bibinfo{author}{M.~Viel},
\newblock \bibinfo{title}{{Tomography of the Fermi-lat $\gamma$-ray Diffuse
  Extragalactic Signal via Cross Correlations With Galaxy Catalogs}},
\newblock \bibinfo{journal}{Astrophys. J. Suppl.} \bibinfo{volume}{217}
  (\bibinfo{year}{2015}) \bibinfo{pages}{15}.
  \DOIprefix\doi{10.1088/0067-0049/217/1/15}.
  \href{http://arxiv.org/abs/1503.05918}{{\tt arXiv:1503.05918}}.
\bibitem[{Tröster et~al.(2016)}]{Troster:2016sgf}
\bibinfo{author}{T.~Tröster}, et~al.,
\newblock \bibinfo{title}{{Cross-correlation of weak lensing and gamma rays:
  implications for the nature of dark matter}}  (\bibinfo{year}{2016}).
  \DOIprefix\doi{10.1093/mnras/stx365}.
  \href{http://arxiv.org/abs/1611.03554}{{\tt arXiv:1611.03554}},
  \bibinfo{note}{[Mon. Not. Roy. Astron. Soc.467,no.3,2706(2017)]}.
\bibitem[{Cuoco et~al.(2017)Cuoco, Bilicki, Xia, and Branchini}]{Cuoco:2017bpv}
\bibinfo{author}{A.~Cuoco}, \bibinfo{author}{M.~Bilicki},
  \bibinfo{author}{J.-Q. Xia}, \bibinfo{author}{E.~Branchini},
\newblock \bibinfo{title}{{Tomographic imaging of the Fermi-LAT gamma-ray sky
  through cross-correlations: A wider and deeper look}}
  (\bibinfo{year}{2017}). \DOIprefix\doi{10.3847/1538-4365/aa8553}.
  \href{http://arxiv.org/abs/1709.01940}{{\tt arXiv:1709.01940}},
  \bibinfo{note}{[Astrophys. J. Suppl.232,10(2017)]}.
\bibitem[{Branchini et~al.(2017)Branchini, Camera, Cuoco, Fornengo, Regis,
  Viel, and Xia}]{Branchini:2016glc}
\bibinfo{author}{E.~Branchini}, \bibinfo{author}{S.~Camera},
  \bibinfo{author}{A.~Cuoco}, \bibinfo{author}{N.~Fornengo},
  \bibinfo{author}{M.~Regis}, \bibinfo{author}{M.~Viel}, \bibinfo{author}{J.-Q.
  Xia},
\newblock \bibinfo{title}{{Cross-correlating the $\gamma$-ray sky with Catalogs
  of Galaxy Clusters}},
\newblock \bibinfo{journal}{Astrophys. J. Suppl.} \bibinfo{volume}{228}
  (\bibinfo{year}{2017}) \bibinfo{pages}{8}.
  \DOIprefix\doi{10.3847/1538-4365/228/1/8}.
  \href{http://arxiv.org/abs/1612.05788}{{\tt arXiv:1612.05788}}.
\bibitem[{Colavincenzo et~al.(2020)Colavincenzo, Tan, Ammazzalorso, Camera,
  Regis, Xia, and Fornengo}]{Colavincenzo:2019jtj}
\bibinfo{author}{M.~Colavincenzo}, \bibinfo{author}{X.~Tan},
  \bibinfo{author}{S.~Ammazzalorso}, \bibinfo{author}{S.~Camera},
  \bibinfo{author}{M.~Regis}, \bibinfo{author}{J.-Q. Xia},
  \bibinfo{author}{N.~Fornengo},
\newblock \bibinfo{title}{{Searching for gamma-ray emission from galaxy
  clusters at low redshift}},
\newblock \bibinfo{journal}{Mon. Not. Roy. Astron. Soc.} \bibinfo{volume}{491}
  (\bibinfo{year}{2020}) \bibinfo{pages}{3225--3244}.
  \DOIprefix\doi{10.1093/mnras/stz3263}.
  \href{http://arxiv.org/abs/1907.05264}{{\tt arXiv:1907.05264}}.
\bibitem[{Tan and Colavincenzo(2019)}]{Tan:2019gmb}
\bibinfo{author}{X.~Tan}, \bibinfo{author}{M.~Colavincenzo},
\newblock \bibinfo{title}{{Bounds on WIMP dark matter from galaxy clusters at
  low redshift}}  (\bibinfo{year}{2019}).
  \href{http://arxiv.org/abs/1907.06905}{{\tt arXiv:1907.06905}}.
\bibitem[{Akrami et~al.(2018)}]{Akrami:2018mcd}
\bibinfo{author}{Y.~Akrami}, et~al. (\bibinfo{collaboration}{Planck}),
\newblock \bibinfo{title}{{Planck 2018 results. IV. Diffuse component
  separation}}  (\bibinfo{year}{2018}).
  \href{http://arxiv.org/abs/1807.06208}{{\tt arXiv:1807.06208}}.
\bibitem[{Kamionkowski(1996)}]{Kamionkowski:1996ra}
\bibinfo{author}{M.~Kamionkowski},
\newblock \bibinfo{title}{{Matter microwave correlations in an open universe}},
\newblock \bibinfo{journal}{Phys. Rev.} \bibinfo{volume}{D54}
  (\bibinfo{year}{1996}) \bibinfo{pages}{4169--4170}.
  \DOIprefix\doi{10.1103/PhysRevD.54.4169}.
  \href{http://arxiv.org/abs/astro-ph/9602150}{{\tt arXiv:astro-ph/9602150}}.
\bibitem[{Song et~al.(2007)Song, Hu, and Sawicki}]{Song:2006ej}
\bibinfo{author}{Y.-S. Song}, \bibinfo{author}{W.~Hu},
  \bibinfo{author}{I.~Sawicki},
\newblock \bibinfo{title}{{The Large Scale Structure of f(R) Gravity}},
\newblock \bibinfo{journal}{Phys. Rev.} \bibinfo{volume}{D75}
  (\bibinfo{year}{2007}) \bibinfo{pages}{044004}.
  \DOIprefix\doi{10.1103/PhysRevD.75.044004}.
  \href{http://arxiv.org/abs/astro-ph/0610532}{{\tt arXiv:astro-ph/0610532}}.
\bibitem[{Limber(1954)}]{Limber:1954zz}
\bibinfo{author}{D.~N. Limber},
\newblock \bibinfo{title}{{The Analysis of Counts of the Extragalactic Nebulae
  in Terms of a Fluctuating Density Field. II}},
\newblock \bibinfo{journal}{Astrophys. J.} \bibinfo{volume}{119}
  (\bibinfo{year}{1954}) \bibinfo{pages}{655}. \DOIprefix\doi{10.1086/145870}.
\bibitem[{Szapudi et~al.(2001)Szapudi, Prunet, and Colombi}]{Szapudi:2001qj}
\bibinfo{author}{I.~Szapudi}, \bibinfo{author}{S.~Prunet},
  \bibinfo{author}{S.~Colombi},
\newblock \bibinfo{title}{{Fast clustering analysis of inhomogeneous megapixel
  cmb maps}}  (\bibinfo{year}{2001}).
  \href{http://arxiv.org/abs/astro-ph/0107383}{{\tt arXiv:astro-ph/0107383}}.
\bibitem[{Chon et~al.(2004)Chon, Challinor, Prunet, Hivon, and
  Szapudi}]{Chon:2003gx}
\bibinfo{author}{G.~Chon}, \bibinfo{author}{A.~Challinor},
  \bibinfo{author}{S.~Prunet}, \bibinfo{author}{E.~Hivon},
  \bibinfo{author}{I.~Szapudi},
\newblock \bibinfo{title}{{Fast estimation of polarization power spectra using
  correlation functions}},
\newblock \bibinfo{journal}{Mon. Not. Roy. Astron. Soc.} \bibinfo{volume}{350}
  (\bibinfo{year}{2004}) \bibinfo{pages}{914}.
  \DOIprefix\doi{10.1111/j.1365-2966.2004.07737.x}.
  \href{http://arxiv.org/abs/astro-ph/0303414}{{\tt arXiv:astro-ph/0303414}}.
\bibitem[{Efstathiou(2004)}]{Efstathiou:2003tv}
\bibinfo{author}{G.~Efstathiou},
\newblock \bibinfo{title}{{A Maximum likelihood analysis of the low CMB
  multipoles from WMAP}},
\newblock \bibinfo{journal}{Mon. Not. Roy. Astron. Soc.} \bibinfo{volume}{348}
  (\bibinfo{year}{2004}) \bibinfo{pages}{885}.
  \DOIprefix\doi{10.1111/j.1365-2966.2004.07409.x}.
  \href{http://arxiv.org/abs/astro-ph/0310207}{{\tt arXiv:astro-ph/0310207}}.
\bibitem[{Challinor and Chon(2005)}]{Challinor:2004pr}
\bibinfo{author}{A.~Challinor}, \bibinfo{author}{G.~Chon},
\newblock \bibinfo{title}{{Error analysis of quadratic power spectrum estimates
  for CMB polarization: Sampling covariance}},
\newblock \bibinfo{journal}{Mon. Not. Roy. Astron. Soc.} \bibinfo{volume}{360}
  (\bibinfo{year}{2005}) \bibinfo{pages}{509--532}.
  \DOIprefix\doi{10.1111/j.1365-2966.2005.09076.x}.
  \href{http://arxiv.org/abs/astro-ph/0410097}{{\tt arXiv:astro-ph/0410097}}.
\bibitem[{Landy and Szalay(1993)}]{Landy:1993yu}
\bibinfo{author}{S.~D. Landy}, \bibinfo{author}{A.~S. Szalay},
\newblock \bibinfo{title}{{Bias and variance of angular correlation
  functions}},
\newblock \bibinfo{journal}{Astrophys. J.} \bibinfo{volume}{412}
  (\bibinfo{year}{1993}) \bibinfo{pages}{64}. \DOIprefix\doi{10.1086/172900}.
\bibitem[{Ammazzalorso et~al.(2018)Ammazzalorso, Fornengo, Horiuchi, and
  Regis}]{Ammazzalorso:2018evf}
\bibinfo{author}{S.~Ammazzalorso}, \bibinfo{author}{N.~Fornengo},
  \bibinfo{author}{S.~Horiuchi}, \bibinfo{author}{M.~Regis},
\newblock \bibinfo{title}{{Characterizing the local gamma-ray Universe via
  angular cross-correlations}},
\newblock \bibinfo{journal}{Phys. Rev.} \bibinfo{volume}{D98}
  (\bibinfo{year}{2018}) \bibinfo{pages}{103007}.
  \DOIprefix\doi{10.1103/PhysRevD.98.103007}.
  \href{http://arxiv.org/abs/1808.09225}{{\tt arXiv:1808.09225}}.
\bibitem[{Ackermann et~al.(2018)}]{Ackermann:2018wlo}
\bibinfo{author}{M.~Ackermann}, et~al. (\bibinfo{collaboration}{Fermi-LAT}),
\newblock \bibinfo{title}{{Unresolved Gamma-Ray Sky through its Angular Power
  Spectrum}},
\newblock \bibinfo{journal}{Phys. Rev. Lett.} \bibinfo{volume}{121}
  (\bibinfo{year}{2018}) \bibinfo{pages}{241101}.
  \DOIprefix\doi{10.1103/PhysRevLett.121.241101}.
  \href{http://arxiv.org/abs/1812.02079}{{\tt arXiv:1812.02079}}.

\end{thebibliography}

\end{document}